%% file: main.tex
\documentclass[subscriptcorrection,upint,varvw,mathalfa=cal=cm,balance,hyphenate,french,pdf-a, nofoot, nolists]{asmejour} %

\usepackage{import}
\usepackage[frozencache=true,cachedir=minted-cache]{minted}
\usepackage[symbol]{footmisc}
\usepackage{amsfonts}
\usepackage{bbm}
\newtheorem{definition}{Definition}[section]
\usepackage{amsmath}
\usepackage{multirow}
\DeclareMathOperator*{\argmax}{arg\,max}

\hypersetup{%
	pdfauthor={Thomas N. Cintra},                       		   	
	pdftitle={Curve LP Metrics Analysis},                  	
	pdfkeywords={Liquidity provision, Curve Finance, DEX, DeFi, Stablecoins, AMM}, 
	pdfsubject = {Analysis of Curve LP metrics developed by Xenophon Labs.},			
}
\makeatletter
\def\blfootnote{\gdef\@thefnmark{}\@footnotetext}
\makeatother
\titleformat{\subsubsection}[runin]{\itshape}{\thesubsubsection}{1em}{}

\begin{document}

    
    \SetAuthorBlock{Thomas Cintra}{
        Xenophon Labs\\
        thomas@xenophonlabs.com
    } 
    
    
    \SetAuthorBlock{Max Holloway}{%
        Xenophon Labs \\
        max@xenophonlabs.com
    }
    
    \title{Detecting Depegs: Towards Safer Passive Liquidity Provision on Curve Finance}


    \begin{abstract}
        We consider a liquidity provider's (LP's) exposure to stablecoin and liquid staking derivative (LSD) depegs on Curve's StableSwap pools. We construct a suite of metrics designed to detect potential asset depegs based on price and trading data. Using our metrics, we fine-tune a Bayesian Online Changepoint Detection (\textsc{BOCD}) algorithm to alert LPs of potential depegs before or as they occur. We train and test our changepoint detection algorithm against Curve LP token prices for 13 StableSwap pools throughout 2022 and 2023, focusing on relevant stablecoin and LSD depegs. We show that our model, trained on 2022 UST data, is able to detect the USDC depeg in March of 2023 at 9pm UTC on March 10th, approximately $5$ hours before USDC dips below $99$ cents, with few false alarms in the 17 months on which it is tested. Finally, we describe how this research may be used by Curve's liquidity providers, and how it may be extended to dynamically de-risk Curve pools by modifying parameters in anticipation of potential depegs. This research underpins an API developed to alert Curve LPs, in real-time, when their positions might be at risk.
    \end{abstract} 

    \date{}
    
    \maketitle 

    \blfootnote{This research was sponsored by Llama Risk and the Curve Research team. Thank you to Benny Lada, WormholeOracle, Nagaking, and Chan-Ho Suh for their continued support.}
    


    \import{./sections}{introduction.tex}
    \import{./sections}{metrics.tex}
    \import{./sections}{detection.tex}
    \import{./sections}{benchmark.tex}
    \import{./sections}{results.tex}
    \import{./sections}{takeaways.tex}

    \clearpage

    \appendix   

    \onecolumn

    \import{./appendix}{bocd_setup.tex}

    \clearpage

    \import{./appendix}{pools.tex}

    \clearpage

    \import{./appendix}{sources.tex}

    \clearpage

    \import{./appendix}{fpr.tex}

    \clearpage
    
    \bibliographystyle{asmejour}   
    \bibliography{bibliography} 
    
    
    
\end{document}

%% file: sections/introduction.tex
\section{Introduction}

    In this paper we provide promising evidence for how quantitative metrics and detection algorithms can be constructed to keep liquidity providers (LPs) informed, in real-time, regarding potential stablecoin and liquid staking derivative depegs. We first briefly introduce what stablecoins are, how they are traded on Curve, and the risks faced by liquidity providers (LPs). A reader familiar with these concepts may skip to Section \ref{subsec:paper}.

    \subsection{Background}

        Stablecoins and liquid staking derivatives are tokens pegged to some underlying floating currency, referred to as a numeraire. Stablecoins are generally pegged to the U.S. Dollar, whereas liquid staking derivatives are generally pegged to ETH or other network tokens, like SOL. Organizations issuing pegged tokens maintain a promise to redeem each token for exactly one of the numeraire asset. Often, they maintain this promise by holding reserves of the numeraire that match or exceed the outstanding supply of the pegged asset. However, if traders lose faith in the token's redemption mechanism they might begin to sell the token, causing prices to decline in secondary markets. We refer to this process as a depeg. Depegs may be momentary, meaning that the token prices on secondary markets eventually return to their peg, or the depeg may be permanent, meaning they never return to their peg. The canonical case studies for depegs in DeFi are the collapse of the Terra Luna ecosystem in May of 2022, the momentary depeg of USDC during the collapse of Silicon Valley Bank in March of 2023, and the multiple momentary depegs, followed by the permanent depeg of Neutrino USD (USDN) in 2022. These three stablecoin depegs will be the primary case studies of this paper, along with the prolonged momentary depeg of stETH from ETH throughout 2022.
    
        Note that currency depegs are not endemic to crypto markets, a classic example of a depeg is the Thai Baht depeg in 1997 when the Thai monetary authority decided to float its currency relative to the U.S. Dollar. For an introduction to currency depegs in both fiat and DeFi, refer to \textit{When the Currency Breaks} by Michael Bolger and Henry Hon \cite{bolgerresearch}.
    
        Curve Finance has pioneered a new decentralized financial protocol (DeFi protocol) for minimizing slippage when swapping pegged assets, particularly stablecoins. Curve is an automated market maker (AMM) that allows traders to swap tokens along some price curve determined by the relative liquidity of tokens within the market - called a pool. For Curve, the price function is a combination of a constant product and a constant sum function, which allows traders to swap with lower slippage when prices deviate slightly from their pegs, as compared to a pure constant price AMM like Uniswap v2 \cite{egorov2019stableswap}. This makes Curve a very attractive venue for traders to swap stablecoins, and therefore an attractive venue for providing liquidity and earning a portion of fees generated.
    
\begin{figure}
            \centering\includegraphics[width=\linewidth]{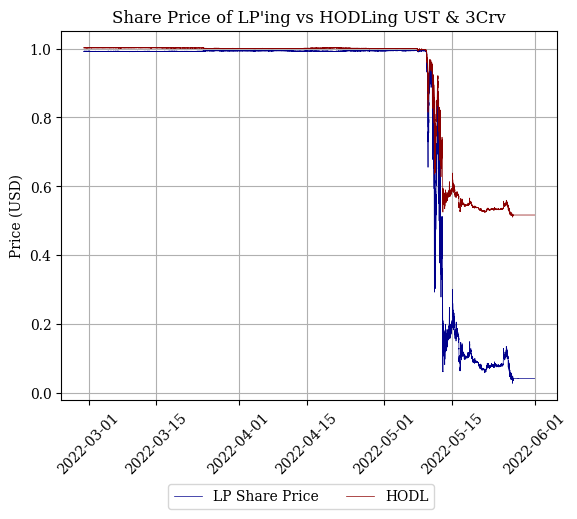}
            \caption{An illustration of impermanent loss for LPs in Curve's UST-3Crv MetaPool during the UST crash. We compare the price of 1 LP share (or LP token) for the UST-3Crv pool to the ``price'' of a portfolio composed of equal parts UST and 3Crv. We assume a price of 1 for 3Crv (that is, we ignore fees accrued) and we benchmark the UST-3Crv LP share price using the pool's real-time reserves and centralized exchange prices from \texttt{ccxt}. \label{fig:imploss}}
        \end{figure}

        An LP on Curve is exposed to two primary risks: (1) they are exposed to the market risk of any of their tokens depegging, (2) they are exposed to the impermanent loss they'd suffer in the pool if a token did depeg. Figure \ref{fig:imploss} illustrates the distinction between these two risks by comparing the USD value of an LP share in Curve's UST-3Crv MetaPool to the USD value of a balanced portfolio holding UST and the 3Crv token. 
    
        Impermanent loss is a comparison of two strategies - the LP's strategy and a ``HODL'' strategy - for allocating a portfolio $\{x_i\}$: the LP's strategy is to deposit into an AMM and earn a fee income, the HODLer's strategy is to simply hold [equal parts of] each token \cite{pintail2019uniswap}. As prices fluctuate in favor or against the portfolio, the LP strictly loses value compared to HODL if we disregard fees. For example: suppose informed traders are confident the price of a token will decline and they begin selling the token in an AMM. On the receiving end of that trade is an LP who is passively buying the declining token. The LP now holds more of the lower-price token than their original portfolio $\{x_i\}$ and the amount by which they lost value compared to the HODL portfolio is the LP's impermanent loss. Of course, the LP might in fact have profited in this scenario if their fee income sufficiently offsets their impermanent loss. On Curve in particular, where stablecoin and LSD prices often revert back to their mean (or pegged) prices, the LP is unlikely to ever realize their impermanent loss, and in fact just profits off the fees.
        
        However, in the case of a depeg, LPs are very exposed to these losses. Specifically, they exhibit very strong negative convexity \cite{aigner2021}. As a token depegs, and the constant-product element of Curve's liquidity-price curve kicks in, the LP begins to buy a lot of the depegging token, and sells what they have of the remaining tokens. In the context of derivatives trading, LPs hold a short straddle position on each of the pool's tokens: they are betting prices will not persistently move in either direction \cite{aigner2021, milionis2022automated}. Find a formal derivation of the StableSwap invariant's impermanent loss here \cite{tiruviluamala2022general}.
    
    
    \subsection{This Paper} \label{subsec:paper} The primary contribution of this paper is to identify strategies for alerting LPs that their positions might be at risk. To the best of our knowledge, this is the first paper that aims to develop online strategies for detecting stablecoin and LSD depegs. We do not claim to predict whether LP'ing during a potential depeg will be a lucrative investment. Intuitively, momentary depegs generate significant increases in swapping activity, resulting in significant boosts to an LP's fee revenue. This increase in profits comes with the increase in risk of prices not mean-reverting. 

    Our strategies are based on changepoint detection algorithms, sometimes referred to as regime shift detection algorithms in biological and environmental science literature. In Section \ref{sec:metrics} we present our metrics, in Section \ref{sec:detection} we describe our chosen changepoint detection strategy, in Section \ref{sec:benchmark} we formalize the scoring rule for our strategy, and we present our results in Section \ref{sec:results}. We develop changepoint detection algorithms using our metrics for two primary reasons:

    \begin{enumerate}
        \item Using these algorithms, we may determine which of our metrics are useful in detecting depegs ahead of time.
        \item These algorithms allow us to make decisions using these metrics. That is, they formalize when a new observation $x_t$ is or is not a changepoint.
    \end{enumerate}

    Finally, we discuss the implications of these detection strategies in Section \ref{sec:takeaways}. Particularly, we describe the alerting tool we have developed to alert LPs in real-time that their positions might be at risk. We show that our \textsc{bocd} strategy identifies certain major depegs, such as the USDC depeg in March of 2022, approximtely 5 hours in advance of USDC dipping below $99$ cents. We propose how this research may be extended to re-calibrate StableSwap pool parameters using the \href{https://github.com/curveresearch/curvesim}{\texttt{curvesim}} package during moments of increased volatility. We discuss how dynamically recalibrating pool parameters could provide a significant improvement in de-risking Curve pools for liquidity providers, and why this dynamic recalibration is itself a difficult problem, particularly in terms of governance.

    Our models have been deployed, and one may listen to their alerts by following \href{https://twitter.com/curvelpmetrics}{this} Twitter account.

    \subsection{Our API} This research paper is the theoretical underpinning for an API developed by the authors for the Curve Research team. You can find our repository \href{https://github.com/xenophonlabs/curve-lp-metrics/tree/main}{here}.

    \subsection{Changepoint Detection} Changepoint detection methods have been developed for several purposes, ranging from marine and environmental science research \cite{rodionov2005brief}, to algorithmic trading strategies \cite{chen2020detecting} and market microstructure analysis \cite{jrfm13100226}. Several studies have looked to review the plethora of changepoint detection strategies \cite{li2019review, rodionov2005brief, vandenburg2020evaluation}, and benchmark them to to real world datasets \cite{vandenburg2020evaluation}. Here, we briefly describe the different kinds of change point detection algorithms. 

    Changepoint detection methods are often performed ``offline'', meaning changepoints are identified in hindsight once all the relevant data has been collected. Furthermore, research is often centered around detecting abrupt shifts in the mean of a timeseries. That is, the underlying data distribution has a constant variance, but undergoes a stepwise change in its mean at some unkown time. Often, these detection algorithms are parametric, meaning we assume some underlying probability density function for our data, and determine when the parameters of this distribution changes. Equation \ref{eq:gaussian_changepoint} formalizes the Gaussian example for the constant-variance, abruptly-changing mean scenario, where the detection algorithm aims to determine the changepoint time $\tau$.

    \begin{equation} \label{eq:gaussian_changepoint}
        X_t \sim
        \begin{cases} 
            \mathcal{N}(\mu_1, \sigma^2), t < \tau \\
            \mathcal{N}(\mu_2, \sigma^2), t \geq \tau
        \end{cases}
    \end{equation}

    As we will discuss in Section \ref{sec:detection}, we test for a variation of Eq. \ref{eq:gaussian_changepoint} using a Student's t-distribution with changing parameters $(\nu, \sigma^2, \mu)$. Importantly, we need our detection algorithm to operate in an ``online'' capacity, meaning each new data point must be analyzed as a potential change point as the data is ingested in real-time. This online requirement rules out a significant portion of standard detection algorithms. 

    In our literature review, we find that the most practical algorithm for change point detection satisfying our requirements is the Bayesian Online Changepoint Detection (\textsc{bocd}) algorithm proposed by Adams and MacKay in 2007 \cite{adams2007bayesian, vandenburg2020evaluation}. We overview the theoretical foundation of \textsc{bocd} and our implementation in Section \ref{sec:detection} and analyze its results on our dataset in Section \ref{sec:results}.

    Before landing on Bayesian Online Changepoint Detection we considered a number of alternative changepoint detection algorithms. These included simpler approaches using Bollinger bands, and cumulative sum (CUSUM) detection, as well as more sophisticated Markov Switching models. We found that \textsc{bocd} was the best suited model for online detection we considered as it is conceptually simple and straightforward to implement.

%% file: sections/metrics.tex
\section{Metrics Overview} \label{sec:metrics}

    We have developed several metrics for detecting depegs for Curve's StableSwap pool. The metrics are summarized in Table \ref{table:metrics}. With each metric we describe the intuition for why this metric might be a useful indicator of potential depegs. We look to identify changes in market behavior that precede changes in token prices.
    
    \begin{table}[htp]
        \centering
        \caption{Summary of Studied Metrics.\label{table:metrics}}
        \begin{tabular}{|>{\raggedright\arraybackslash}p{3cm} >{\raggedright\arraybackslash\hangindent=1em}p{3cm}|}
            \toprule
            Metric & Description \\ 
            \midrule
            Gini Coefficient and Shannon's Entropy & A measurement of the relative balances of a pool's tokens. \\ 
            \hline
            Net Swap Flows & The net amount swapped into or out of a pool for a particular token. \\
            \hline
            Net LP Flows &  The net amount deposited into or withdrawn from a pool for a particular token. \\
            \hline
            Price Volatility & The rolling log returns for a token's price. \\
            \hline
            PIN & The probability of informed trading developed by Easley et al. \cite{easley1996}. \\
            \hline
            Markouts & A short-term measurement of a trade's profits. \\
            \hline
            Shark Trades & A classification of traders/LPs as ``sharks'' based on their past performance. \\
            \bottomrule
        \end{tabular}
    \end{table}

    \subsection{Gini Coefficient and Shannon's Entropy}

        We begin with two metrics meant to capture changes in the token balances of each pool. Intuitively, abrupt changes in the relative token balances in the pool could signal new information regarding one or more of the underlying tokens. For example, informed traders might sell the depegging token to the pool in large amounts. The fire sale of the depegging token might be abrupt, as with UST, or gradual, as with USDN. Measuring changes in the relative weights of each token in the pool might therefore be a useful indicator of potential depegs. 
        
        We consider two related metrics for measuring relative balances of pools with $2$ or more tokens: the Gini Coefficient and Shannon's Entropy. Both metrics allow us to produce a scalar value based on pool composition for pools with an arbitrary number of tokens. Ultimately, we find that Shannon's entropy produces a useful signal for detecting depegs.

        \subsubsection{The Gini Coefficient} is a measure of inequality, originally developed to quantify income or wealth distribution within a population. The Gini coefficient ranges from 0 (representing perfect equality, where the observed Lorenz Curve matches the Line of Perfect Equality) to 1 (representing perfect inequality). For a discrete probability distribution $p(x)$ of $N$ values sorted in non-decreasing order, the Gini coefficient $G$ can be expressed as:
        
        \begin{equation}
            G = \frac{1}{N-1} \sum_{i=1}^{N} (2i - N - 1)p(x_i)
        \end{equation}
        
        where $x_i$ is the $i^{th}$ value in the sorted list, and the sum runs over all values.
        
        The Gini coefficient has been widely adopted in various fields due to its interpretative simplicity and its sensitivity to the changes at the extremes of distributions. In the context of Curve's StableSwap pools, the Gini Coefficient is calculated on the token balances of the pool. A perfectly balanced pool, where each token has an equal share of the pool's total balance, would have a Gini Coefficient of zero. 

        \subsubsection{Shannon's Entropy} is a fundamental concept in information theory, introduced by Claude E. Shannon in his seminal 1948 paper \textit{A Mathematical Theory of Communication} \cite{shannon2001mathematical}. It quantifies the uncertainty or randomness in a dataset, essentially measuring the informational diversity. In the AMM context: imagine a pool as a basket where each unit of each token is a ball with that token's color. Shannon entropy loosely measures how easy it is to predict the color of a ball picked at random from that basket. It can therefore be conceptualized as a variation of the Urn problem.
        
        For a discrete random variable $X$ with possible values $\{x_1, x_2, \ldots, x_n\}$ and probability mass function $p(x)$, the Shannon entropy $H(X)$ is defined as:
        
        \begin{equation}
            H(X) = -\sum_{i=1}^{n} p(x_i) \log_2 p(x_i)
        \end{equation}
        
        In this equation, $p(x_i)$ is the probability of event $x_i$ occurring. The logarithm base is typically 2 if the measure of entropy is in bits, but any base can be used. The entropy is maximized when all outcomes are equally likely, indicating maximum uncertainty or randomness.

\begin{figure}[htp]
            \centering\includegraphics[width=\linewidth]{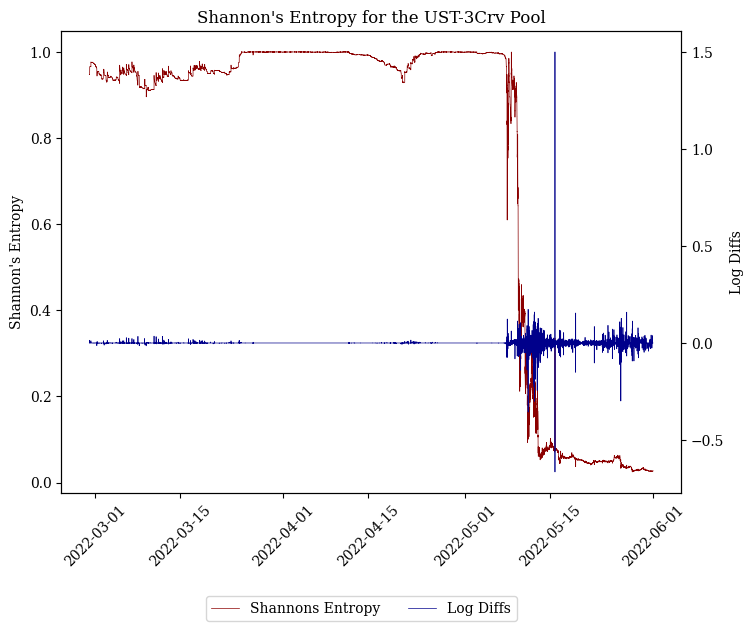}
            \caption{Shannon's entropy calculated on a minutely basis for the UST-3Crv pool. The log diffs are calculated as $\log{(1 + \frac{x_{t+1}}{x_t})}$. Notice that as entropy decreases, smaller absolute changes in entropy lead to much larger percentage changes, which is why we observe a large spike on May 15th. \label{fig:shannon}}
        \end{figure}

        We may produce a signal with either the Gini Coefficient or Shannon's Entropy by continuously measuring the diffs (or log diffs) on a periodic (per block, or per minute) basis. Ideally, our changepoint detection algorithm would signal significant changes in the variance of these returns, as shown in Fig. \ref{fig:shannon}. We take the logarithmic difference of our metrics in an attempt to create a stationary timeseries, and mitigate the noise in their variances.

        We will refer to Shannon's entropy as \textit{entropy} for the remainder of this paper.

    \subsection{Net Swap Flows}

        Whenever a swap is conducted in an AMM it moves the relative prices of the swapped tokens. This creates an arbitrage opportunity for traders to buy low and sell high relative to other markets. This means that over sufficiently long windows of time, denoted as $\Delta t$, trades will be executed to rebalance AMM prices immediately following a swap (i.e. backrunning). It follows that the net swap flow for the pool over $\Delta t$ is zero in expectation, if token prices stay approximately constant. For example, if USDC/USDT prices are approximately constant around $1$, we would expect the net amount of USDC or USDT being swapped into or out of Curve's 3pool to be approximately zero for both tokens. This maintains the relative prices of USDC and USDT within the 3pool approximately constant at $1$\footnote{This assumes balanced withdrawals by LPs. If LPs withdraw significant amounts of just one token, arbitrageurs will be incentivized to sell that token to the pool, and we will observe a net positive swap flow.}. 

        However, if informed traders begin to price a token below its peg, it is very likely that we will observe non-zero net swap flows. Since takers are generally more informed than makers, large non-zero net swap flows might be highly predictive of a change in a token's price. Ideally, this metric would allow us to detect a potential depeg before this information dissipates, and the market prices in the new price of the token. That is, the increase or decrease in net swap flows would ideally precede the change in the market price of a token. This is might be especially true if we filter the net swap flows to only track informed traders. We describe an attempt in doing so with our ``sharks'' metric.

        Formally, we define the net swap flow over rolling window $\Delta t$ for token $i$ on pool $P$ as:

        \begin{multline} \label{eq:netSwapFlow}
            \text{netSwapFlow}(P, i \Delta t) = \nonumber \text{Token $i$ buy volume on pool $P$ over $\Delta t$}\\ - \text{Token $i$ sell volume on pool $P$ over $\Delta t$}
        \end{multline}

        \begin{figure}[htp]
            \centering\includegraphics[width=\linewidth]{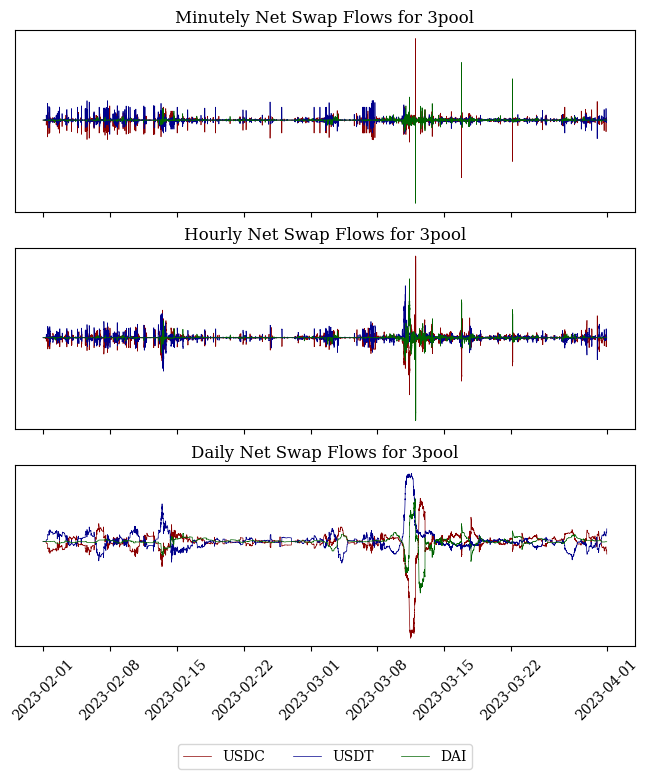}
            \caption{Net swap flows for different rolling window sizes in Curve's 3pool during the collapse of Silicon Valley Bank. The mean in each plot is zero. \label{fig:swapflows}}
        \end{figure}

        Figure \ref{fig:swapflows} shows a snapshot of our net swap flow metric for Curve's 3pool during the collapse of Silicon Valley Bank. Notice that swap flows usually fluctuate around 0 as expected, but with non-constant variance.

    \subsection{Net Deposit and Withdrawal Flows}

        Similarly to net swap flows, we measure the rolling window of net deposits and withdrawals into the pool. We denote this as the net LP flow, defined very similarly for each token as in Eq. \ref{eq:netSwapFlow}\footnote{Note that the deposited token for a MetaPool is the LP token of its base pool. That is, the deposited tokens for the UST-3Crv MetaPool are UST and 3Crv (the LP token for the 3pool), instead of UST, USDC, USDT, and DAI.}.

        We posit that many of the large liquidity providers on Curve are actively and systematically providing liquidity to particular pools. A sudden change in the net deposit and withdrawal flows for a pool might be indicative that these LPs are anticipating an increase in risk - such as a depeg. Tracking this metric would allow all LPs to ``copy trade'' the first-movers in a particular pool.
    
    \subsection{Price Volatility} 
    
        Intuitively the price for each token would be a reasonable indicator of a potential depeg. However, looking directly at the price of a token could not produce a leading indicator in itself. Instead, we consider the volatility of returns for a particular token, and test whether abrupt changes in the token's volatility precede its eventual decline in price. 
        
        For some window $\Delta t$, and prices $p_{i, t}$, the rolling price volatility for token $i$ is:
    
        \begin{align}
            R(i, t_0, t_1) = \left\{R_{i, t \in [t_0, t_1]}\} = \{\ln\left(\frac{p_{i, t+1}}{p_{i, t}}\right)\right\} \\
            \text{volatility} = \sigma_{R}\\
            \nonumber \Delta t = t_1 - t_0
        \end{align}
    
        Where $\sigma_R$ is the standard deviation of $R(i, t_0, t_1)$.

    \subsection{PIN}
    
        The probability of informed trading (PIN) is a metric proposed by Easley et al. in 1996 \cite{easley1996} that measures information asymmetry in a market. The idea is that in a given window of time, traditionally a trading day, an imbalance in buy and sell orders indicates an increased probability of an information event. For example, if a trading day exhibits significantly more buy orders than sell orders compared to previous trading days, it is likely that good news for the underlying asset arrived before or during the trading day, prompting informed selling \cite{foucault2013market}. In theory, a sudden increase in PIN suggests an increase in informed buying or selling, which could then be a useful predictor of a potential depeg. 

        Let $\alpha$ denote the probability that an information event has occured (such as a depeg), and let $\theta$ define the probability that this news was positive. For our purposes, it is likely that $\theta$ is very close to zero (i.e. informed selling or shorting, as opposed to informed buying). The authors posit that informed trades arrive as a Poisson process with intensity $\epsilon_i$, whereas uninformed buys arrive with intensity $\epsilon_b$ and uninformed sells arrive with intensity $\epsilon_s$. As these three Poisson point processes are independent, the likelihood of any trade being informed can be shown to be:

        \begin{equation} \label{eq:pin}
            \text{PIN} = \frac{\alpha \epsilon_i}{\epsilon_b + \epsilon_s + \alpha \epsilon_i}.
        \end{equation}

        These parameters are often estimated with maximum likelihood estimators. The following is the likelihood function for $B$ buy and $S$ sell orders in one trading day \cite{easley2010factoring}, assuming orders are sampled i.i.d. from their respective Poisson point processes:

        \begin{multline}
            \mathcal{L}((B, S)|\epsilon_i, \epsilon_b, \epsilon_s, \alpha, \theta) = \alpha (1 - \theta)e^{-(\epsilon_i + \epsilon_b + \epsilon_s)}\frac{(\epsilon_i + \epsilon_b)^B\epsilon_s^S}{B!S!} \\ + \alpha \theta e^{-(\epsilon_i + \epsilon_b + \epsilon_s} \frac{(\epsilon_i + \epsilon_s)^S\epsilon_b^B}{B!S!} + (1 - \alpha)e^{(\epsilon_b + \epsilon_s)}\frac{\epsilon_b^B\epsilon_s^S}{B!S!}.
        \end{multline}

        This likelihood function can be generalized for $(B_t, S_t)|_{t=1}^T$ orders in a period $T$ using a log-likelihood function, which is what we implement for our modeling \cite{easley2010factoring}. Once we have estimated the relevant parameters $(\epsilon_i, \epsilon_b, \epsilon_s, \alpha, \theta)$ we may calculate the PIN for the relevant window $T$, giving us our rolling PIN metric. Notice that rolling PIN is a computationally expensive metric to calculate relative to other metrics displayed here, as it requires a several steps of numerical approximation of the parameters until convergence, which must be performed at every time step. 

        \begin{figure}[htp]
            \centering\includegraphics[width=\linewidth]{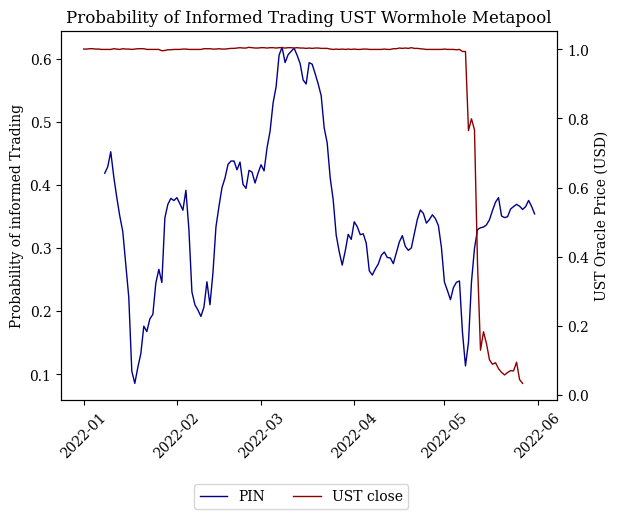}
            \caption{The probability of informed trading fit on weekly rolling window of daily aggregate data for the UST-3Crv pool. Unexpectedly, PIN peaks in March of 2022, and sees only a marginal increase during the depeg in May.\label{fig:pin}}
        \end{figure}

        As shown in Fig. \ref{fig:pin}, we find that PIN does not behave as expected in the depegs we have examined, and is not a useful metric for our purposes.

    \subsection{Short-Term Markouts} 

        Markouts are a common way to evaluate trade performance in the context of high-frequency trading, and have lately been used to evaluate LP performance in AMMs \cite{holloway2022, crocswap2022}. Generally, markouts compare the execution price of a trade to its price at some ``mark'' in the future. Markouts are a powerful way of measuring the ``informedness'' of a trade, meaning that traders with consistently positive markouts for some window $W$ tend to be good at predicting price moves within that time window, such as 5 minutes. Formally, we define the markout $m$ for trade $\phi = (x_i, x_j)$ as

        \begin{equation} \label{eq:markouts}
            m_{\phi} = x_i \cdot \left(P_{t+h} - \bar{p}(\phi) \right),
        \end{equation}

        where $x_i$ quantity of token $i$ is being bought, $P_{t+h}$ is the relative price of tokens $i, j$ at time $t+h$, $h$ being the markout window, and $\bar{p}(\phi)=\frac{x_i}{x_j}$ is the trade's execution price (accounting for fees)\footnote{We have defined markouts from the perspective of takers, we may similarly define them for LPs by multiplying $m_{\phi}$ by negative one (i.e. the LP is selling token $i$).}. Given that Curve pools might have multiple tokens, we normalize the units of our markout by taking the numeraire prices of each token.

        \begin{equation} \label{eq:markouts_dollar_eg}
            m_{\phi} = x_i \cdot p_{i, t+h} - x_j \cdot p_{j, t},
        \end{equation}

        where $p_{i, t}$ is the numeraire price (e.g. USD price) of token $i$ at time $t$. By considering the numeraire-price of each token we it becomes possible for both the taker and the LP to profit off the same trade if the numeraire-price of both tokens increase. However, as we will discuss in the following subsection, we would like our markout metric to clearly determine whether the taker or the LP were more informed at the time of the trade, by assigning the ``less informed'' side a negative markout. We modify the markout metric to allow us to compare the numeraire-price of each token at the mark time.

        \begin{equation} \label{eq:markouts_dollar}
            m_{\phi} = x_i \cdot p_{i, t+h} - x_j \cdot p_{j, t+h}.
        \end{equation}

\begin{figure}[htp]
            \centering\includegraphics[width=\linewidth]{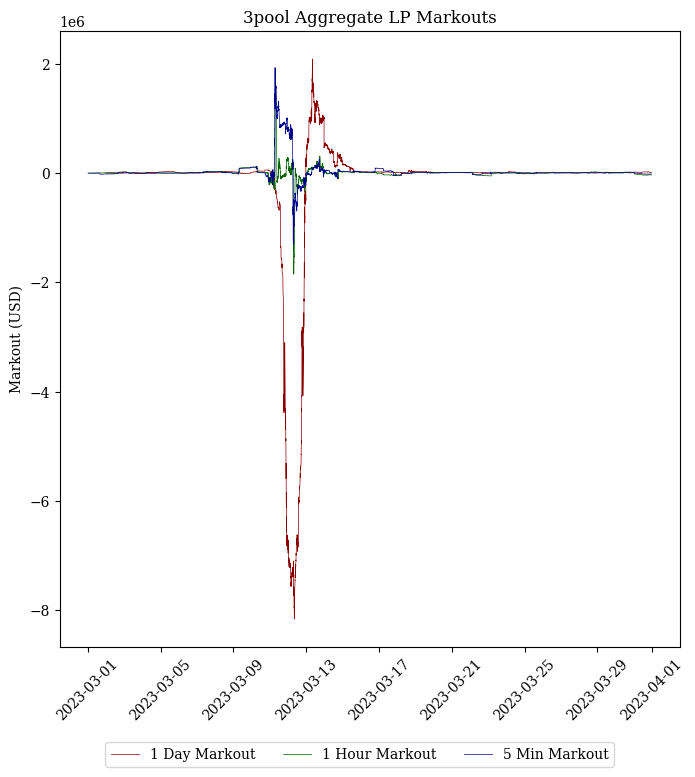}
            \caption{LP markouts for various windows. \label{fig:markouts}}
        \end{figure}

        This allows us to parse through all swaps and deposits/withdrawals made on Curve's major StableSwap pools and rank the users who have historically taken the most advantage of token depegs by selling the depegging token for a pegged token. We expect markouts over time to have a mean around 0 for pegged assets with respect to their numeraires. Large fluctuations in markouts, resulting in changes to the variance of the markouts timeseries, may then be used to alert LPs of potential depegs. As a depeg is about to occur, we expect informed traders to place larger-than-usual bets against the depegging token, which would lead to sudden spikes in markouts that are highly unlikely unless prices begin to depeg, as shown in Fig. \ref{fig:markouts}. While markouts can be leading indicators by providing us with very large signals despite relatively small deviations in price, they must necessarily follow price action, instead of precede it.
        
        Note that markouts are very sensitive to the markout window being considered. Since we are not interested in measuring the LP's or taker's performance \textit{per se} with markouts, we simply look to construct a useful signal for our changepoint detection. It is then better to observe shorter markout windows, since we are able to observe abrupt changes in markouts more quickly.

    \subsection{Sharks}

        As described in the previous subsection, we use our markouts metric to identify informed takers. We define these takers, called ``sharks'', as the takers with the top $1\%$ cumulative markouts. We measure cumulative markouts using all trades executed on the pools considered in our research (refer to Appendix \ref{app:pools}), from January 1st 2022 to May 1st 2023. By specifically looking at previous markout performance, we identify which accounts successfully traded previous depegs in size. 

        Using this classification of sharks, we may then measure the net swap flows for each token based only on shark volume. In theory, this could significantly reduce the amount of noise in our net swap flow metric, creating a more predictive signal. For example, a high net shark volume may indicate sharks have information that the pool does not. 
        
        However, there are a number of complications in measuring shark performance. As previously mentioned, markouts are very sensitive to the markout window; a shark under a 5 minute markout might not be a shark under a 1 day markout and vice-versa. Furthermore, indexing sharks by Ethereum addresses is not ``Sybil resistant'': we are not capturing the cumulative markout of a ``trader'' per se, just one of their addresses. This means, for example, that if a shark trader begins to use a new Ethereum address, we have no way of classifying their new address as a shark address. Finally, markouts are not representative of any other trades or hedges performed on other trading venues, meaning it is not representative of a trader's profit and loss (PnL). Still, we find that 5 minute markouts provide a useful heuristic for the top performing addresses on Curve.

        \begin{figure}[htp]
            \centering\includegraphics[width=\linewidth]{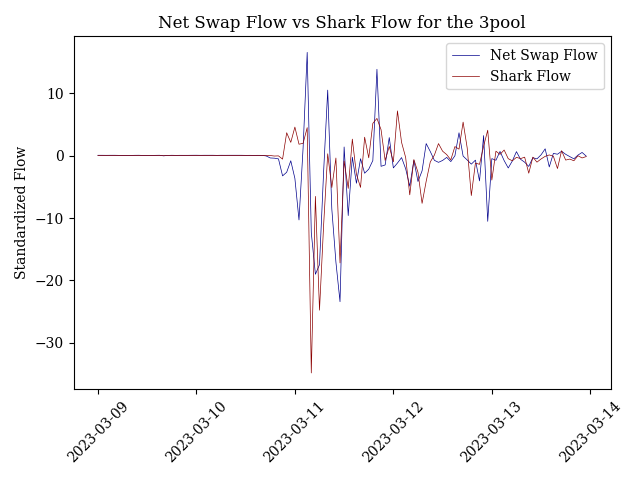}
            \caption{Net shark flow vs net swap flow for the 3pool during the SVB collapse. Notice how these metrics are very closely related, despite us having filtered out $75\%$ of takers in constructing the net shark flow metric. We show these metrics perform similarly in the results section. \label{fig:sharkflow}}
        \end{figure}
    
        For the purposes of classifying sharks that have successfully traded depegs in the past, we employ a 1-day markout (instead of the 5 minute markout) to create our sharkflow metric. 

%% file: sections/detection.tex
\section{Bayesian Online Changepoint Detection} \label{sec:detection}

    Here we provide a brief overview of Bayesian Online Changepoint Detection. An extensive derivation of the formulation for \textsc{bocd} is outside the scope of the paper, refer to the original paper \cite{adams2007bayesian} (particularly Algorithm 1) or other helpful blog posts explaining the intuition and implementation of \textsc{bocd} in Python \cite{gundersen2019, gundersen2020, kompella2020}. 

    Change points are abrupt changes in the generative parameters of some sequence. Formally, denote a sequence of observations as $\vec{x}=\{x_i\}, i \in [1, T]$ where observations are i.i.d. from some probability distribution $P$ with parameters $\eta$: 
    
    \begin{equation}
        X \sim P(x_t | \eta)
    \end{equation}
    
    Suppose our sequence $\vec{x}$ exhibits changes in its underlying distribution, meaning $\eta$ is not constant over time. Suppose there are $n$ changepoints occurring randomly within $[1, T]$. We may then partition our sequence into non-overlapping partitions $\rho = 1, 2, \ldots n$, each with parameters $\eta_{\rho}$. We further denote the contiguous set of observations between $a, b$ as $\vec{x}_{a:b}$.

    In Bayesian online changepoint detection, our goal is to estimate when some observed datum $x_t$ is very unlikely to have been sampled from the current assumed parameters $\eta$. As proposed by Adams and MacKay in their original \textsc{bocd} work \cite{adams2007bayesian}, we do so by tracking a ``run length'' at each data point, denoted as $r_t$. The run length can be understood as the number of time steps since the last change point: with every new datum, $r_t$ either increases by 1, or it drops to $0$.

    At a high level, \textsc{bocd} computes the posterior probability distribution for the run length $r_t$ given the observed data $x_{1:t}$ using Bayes' Theorem:

    \begin{equation}
        P(r_t | x_{1:t}) = \frac{P(r_t, x_{1:t})}{P(x_{1:t})}
    \end{equation}

    where $P(x_{1:t}) = \sum_{r_t}P(r_t, x_{1:t})$. At each time step, we compute $\gamma_t = \argmax{P(r_t | x_{1:t})}$, which denotes the most likely run length for the current partition. Intuitively, a changepoint is identified at $t$ whenever $\gamma_t = 0$. 

    The key insight in \textsc{bocd} is how we may compute $P(r_t, x_{1:t})$ recursively using a specific kind of likelihood function $\pi$, and a hazard function $H$ (described in more detail in Appendix \ref{app:bocd_setup}). We consider two separate cases: $r_t=0$ is a changepoint, or $r_t=l$ is not a changepoint. Notice that $l$ does not necessarily have to be equal to $\gamma_{t-1} + 1$, meaning that a changepoint might have occurred in a previous timestep which was not immediately identified. For the purposes of scoring our \textsc{bocd} strategy, when $\gamma_t = \argmax{P(r_t | x_{1:t})} \neq \gamma_{t-1} + 1$, we shall mark $t$ as a changepoint itself, since that is the earliest step where our \textsc{bocd} strategy identified a discontinuity in run length. Let's consider the joint probability distribution for $r_t$ and $x_{1:t}$ in the case where $r_t=l$:

    \begin{equation} \label{eq:update_l}
        P(r_t=l, x_{1:t}) = P(r_{t-1}, x_{1:t-1}) \cdot \pi_{t-1}^l \cdot (1 - H(r_{t-1}))
    \end{equation}

    where $\pi = P(x_t | \eta)$ is some likelihood function parameterized by $\eta$, and the notation $\pi_{t-1}^l$ denotes the likelihood function with parameters based on $x_{1:t}$ and run length $r_t = l$. For example, we might use a Gaussian distribution such that $\eta = (\mu, \sigma)$, where the mean and standard deviation are updated with every step. In this case, $\pi_{t-1}^l$ denotes the probability that $x_t \sim \mathcal{N}(\mu, \sigma)$ where $\mu, \sigma$ are fit on $x_{t-l}, \ldots x_{t-1}$. 

    The hazard function $H$ simply denotes the probability that the next data point is a changepoint, given a run length $r_{t-1}$ and no additional information. More formally:

    \begin{equation}
        P(r_t | r_{t-1}) =
            \begin{cases} 
            H(r_{t-1}+1), & \text{if } r_t = 0, \\
            1 - H(r_{t-1}+1), & \text{if } r_t = r_{t-1} + 1, \\
            0, & \text{otherwise.}
            \end{cases}
    \end{equation}

    To complement Eq. \ref{eq:update_l}, we compute the probability that the run length has dropped to 0:

    \begin{equation} \label{eq:update_0}
        P(r_t=0, x_{1:t}) = \sum_{r_{t-1}} P(r_{t-1}, x_{1:t-1}) \cdot \pi_{t-1}^l \cdot H(r_{t-1})
    \end{equation}

    In this case we consider all the possible paths $r_{t-1}$ that would lead to a changepoint $r_t=0$. Combining Eqs. \ref{eq:update_l} and \ref{eq:update_0}, we have a recursive formulation for $P(r_t, x_{1:t})$ and we may compute the sequence $\{\gamma_t\}$. Recall that our online algorithm reports changepoints whenever $\gamma_t \neq \gamma_{t-1} + 1$.
    
    In our implementation, we assume a Student's t-distribution for $\pi$ and a constant hazard function $H$. For some further detail on why these choices are sensible, and how we performed hyperparameter tuning, refer to Appendix \ref{app:bocd_setup}.

    \subsection{Example} 
    
        We have described \textsc{bocd} as an exercise in finding the most likely run length $\gamma_t = \argmax{P(r_t | x_{1:t})}$ at every time step, where discontinuities in $\gamma_t$ are emitted as changepoints. Refer back to Fig. \ref{fig:shannon}, specifically to the log differences between each timestep for our entropy measure. For our \textsc{bocd} strategy, we assume that these log differences are sampled from some Student's t-distribution with parameters $\eta = (\nu, \sigma^2, \mu)$. A successful implementation of \textsc{bocd} will hopefully capture the underlying change in the distribution of the log diffs on our entropy metric. For example, we would hopefully capture an increase in the degrees of freedom parameter $\nu$ and/or an increase in the spread (i.e. variance) parameter $\sigma^2$. This would occur when some log diff $x_t$ is observed that is considered unlikely to have been sampled using the previous parameters $\eta$, which would have a very small spread, relative to the magnitude of $x_t$.
    

%% file: sections/benchmark.tex
\section{Benchmarking \textsc{bocd}} \label{sec:benchmark}

    Now that we have formalized our model for changepoint detection, we must define a scoring rule to determine if the model is a successful leading indicator of changepoints for our Curve LPs. Notice that it is not sufficient to evaluate our model against a labelled set of changepoints on our metrics, since the metrics themselves can be noisy. 
    
    Instead, we define a \textsc{baseline depeg model} for identifying depegs, based on the LP token value over time, including fees accrued. Depegs identified by our \textsc{baseline depeg model}, which is also available through our API, will be labeled as ``true changepoints''. We will then use a modified F-score to determine whether our \textsc{bocd} on a particular metric is a leading indicator of potential depegs.

    \subsection{\textsc{baseline depeg model}}
    
        We first define the USD LP share price (or LP token price) for the pool:

        \begin{definition}[LP Share Price]
            We define the LP Share Price as:
            
            \begin{equation} \label{eq:shareprice}
                \text{LP Share Price} = \frac{\langle \vec{x}, \vec{p} \rangle}{\text{LP Token Supply}}
            \end{equation}

            where $\vec{x}$ is the balances of the pool's tokens, $\vec{p}$ is their respective prices, and $\langle ., . \rangle$ is an inner product. Intuitively, the LP Share Price measures how much an LP token is worth, if an LP were to perform a balanced withdrawal using \href{https://github.com/curvefi/curve-contract/blob/b0bbf77f8f93c9c5f4e415bce9cd71f0cdee960e/contracts/pool-templates/base/SwapTemplateBase.vy#L513}{\texttt{remove_liquidity(.)}}, and subsequently sell it instantaneously in an alternative liquid centralized or decentralized exchange (without accounting for slippage)\footnote{Alternatively, Liquidity Provider performance is often measured using markouts; see this analysis of Uniswap v3 markouts by Crocswap \cite{crocswap2022} or our previous work valuing non-toxic orderflow on Uniswap \cite{holloway2022}.}.
        \end{definition}
    
        In the case where liquidity for one or more of the underlying tokens is held primarily on the Curve pool itself, our LP share price might be an overestimation.
        
        Our \textsc{baseline depeg model} will alert LPs of a potential depeg by comparing the LP share price with the virtual price of the LP token. Recall that the virtual price computes the LP token price (including fees accrued), by assuming each underlying token reverts to its peg. Therefore, by comparing the LP share price and the virtual price, we can measure the degree to which changing underlying token prices have affected an LP's portfolio value.

        \begin{definition} [Potential Depeg]
            We define a potential depeg for a Curve pool as any time, $t$, where the LP share price deviates from the virtual price of the pool by $5\%$ or more:

            \begin{equation}
                \frac{\text{Virtual Price}(t) - \text{LP Share Price}(t)}{\text{Virtual Price}(t)} \geq 0.05
            \end{equation}

            Potential depegs will be used as the ``true'' changepoint labels when scoring our metrics. Notice that $5\%$ was chosen to minimize the noisy labels given hourly data, and we could choose a smaller threshold for more coarse data (e.g. $1\%$ for daily data).
        \end{definition}

\begin{figure}[htp]
            \centering\includegraphics[width=\linewidth]{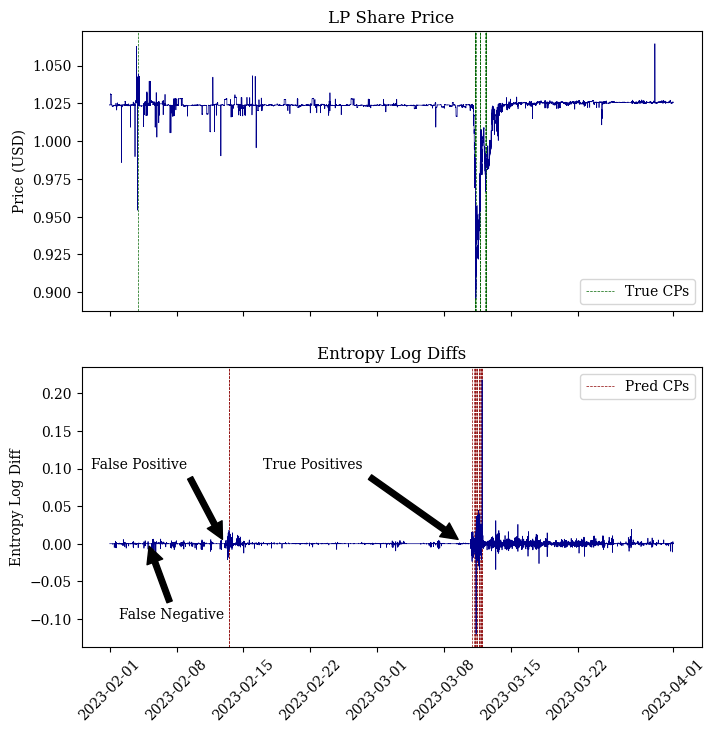}
            \caption{(Top) LP Share Price for 3pool LP tokens during the SVB collapse, as defined by equation \ref{eq:shareprice}. Green dotted lines indicate the potential depegs (i.e. true changepoints) for the pool. (Bottom) Our metric $m$ defined as the log diffs of 3pool entropy, with red dotted lines indicating detected changepoints. The annotations in the plot identify when changepoints should be labelled as a TPs, FPs, TNs, and FNs. Note that we manually set hyperparameters for our Student's t-distribution for the purposes of this illustration, and performance may vary with different parameters.\label{fig:lp_port_entr}}
        \end{figure}

        We may now construct a labelled train and test set to hyperparameter tune our \textsc{bocd} model for each metric. Figure \ref{fig:lp_port_entr} provides an example using the 3pool during the collapse of Silicon Valley Bank and our entropy metric. 
    
        Ultimately, we want to find a set of hyperparameters for our \textsc{bocd} model, such that it predicts depegs using some metric $m$ \textit{before} the LP share price meaningfully decreases. Figure \ref{fig:lp_port_entr} is an example of such a situation, where the first changepoint detected by \textsc{bocd} occurs on March 10th at 9pm GMT, when the LP share price was still $\approx1.0225$. In the following section, we construct an F-score to evaluate a model based on how good of a leading indicator it is for potential depegs.

    \subsection{Scoring Rule}

        We define the following variation of an $F_1$-Score that allows leading indicators to be marked as true positives, modified from \cite{vandenburg2020evaluation}. Furthermore, our $F_{\beta}$-Score uses a weighted Recall measurement, which gives a higher weighting the earlier the changepoint was detected, up to some margin.

        Let $\mathcal{T}$ denote the set of true depegs, calculated from our \textsc{baseline depeg model}, and $\mathcal{X}$ denote the predicted depegs from our \textsc{bocd} model. Both $x \in \mathcal{X}$ and $\tau \in \mathcal{T}$ are in timestamp units. We define the set of true positives as
        
        \begin{equation}\label{eq:true_positives}
            TP(\mathcal{T}, \mathcal{X}) = \left\{\frac{\tau - x}{M} \in \mathcal{T} | \exists x \in \mathcal{X} s.t. 0 \leq \tau - x \leq M\right\}
        \end{equation} 

        
        where $M$ is the maximum leading lag allowed for a predicted depeg $x$ to be considered a leading indicator of $\tau$. That is, the set $TP(\mathcal{T}, \mathcal{X})$ corresponds to all true depegs that were predicted within a leading margin of error $M$, where each TP $x$ is weighed by its distance from the corresponding $\tau$. We may then calculate the precision $P$ and weighted recall $R$:

        \begin{equation}
            P = \frac{|TP(\mathcal{T}, \mathcal{X})|}{|\mathcal{X}|}
        \end{equation}

        \begin{equation}
            R = \frac{\sum{TP(\mathcal{T}, \mathcal{X})}}{|\mathcal{T}|}
        \end{equation}

        Our leading $F_1$-Score measurement may then be defined as follows.

        \begin{definition} [Leading $F_{1}$-Score]
            The leading $F_{1}$-Score measurement for depegs $\mathcal{T}$ and predictions $\mathcal{X}$ with a maximum leading margin $M$ is:

            \begin{equation}
                lF_{1} = \frac{(1+\beta^2)PR}{\beta^2P + R}
            \end{equation}

            where $\beta$ balances the relative importance of $P$ and $R$, with higher $\beta$ biasing in favor of recall. With $\beta=1$, we get the usual $F_1$ score. 
        \end{definition}

        We use this modified $lF_{1}$ scoring rule to evaluate our \textsc{bocd} model in Section \ref{sec:results}, and for hyperparameter tuning in Appendix \ref{app:bocd_setup}.
    

%% file: sections/results.tex
\section{Results} \label{sec:results}

    The results in this section are based on a \textsc{bocd} model trained on hourly UST data from January to June of 2022. We performed a grid search of relevant hyperparameters using our $lF_1$-scoring rule, and tested them on hourly data from January 2022 to May 2023. Hyperparameters were trained and tested for each metric independently, and all training and testing data was standardized\footnote{All our metrics have zero-mean. We standardize metrics because different pools will have different standard deviations (scales/degrees of freedom for t-distribution), but we want to share hyperparameters across pools. By standardizing (meaning we enforce unit-variance), we allow our model to leverage the same hyperparameters for the likelihood distribution across pools.}. Further details regarding hyperparameter tuning are discussed in Appendix \ref{app:bocd_setup}. Our \textsc{bocd} models using these hyperparameters, along with the \textsc{baseline depeg model}, are available through our API. One may listen to their alerts by following \href{https://twitter.com/curvelpmetrics}{this} twitter account and turning notifications on. Figures like \ref{fig:3pool_entropy} can be found for all metrics, pools, and tokens in our GitHub \href{https://github.com/xenophonlabs/curve-lp-metrics/tree/main/figs}{repository}.

    Notice that our \textsc{bocd} model is subject to the usual constraints of statistical or machine learning models. For example, with only a couple years of data and a handful of major depegs to analyze, our models risk overfitting to the variance on the depegs on which it is trained. We chose to train our model on the most relevant training set we have, the UST depeg, and to test it on a wide variety of pools over a long stretch of time. We focus on how our model performed with the momentary depegs of USDC and stETH, as well as the gradual permanent depeg of USDN. Note that hyperparameter tuning is sensitive to training data. Parameters depend particularly on whether the training set contains a relevant depeg. If there is no depeg in the training data, then there is no measurable $lF$-score with which to train hyperparameters. We discuss the implications of this sensitivity in Section \ref{subsec:deployment}.

\begin{figure}
        \centering
        \includegraphics[width=\linewidth]{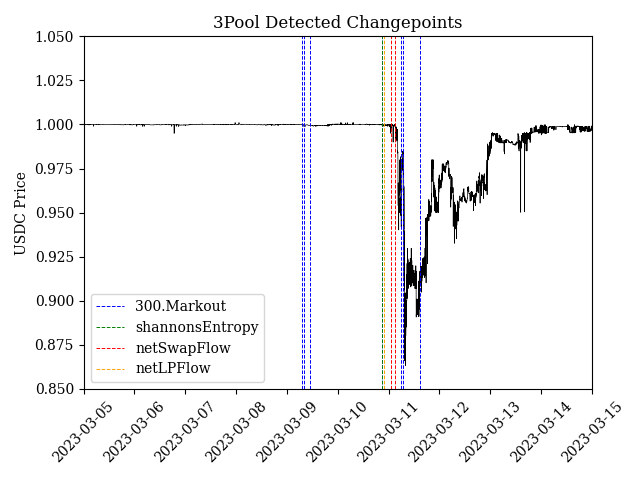}
        \caption{Detected changepoints using various metrics for the 3pool during the collapse of Silicon Valley Bank. The corresponding \textsc{bocd} model was trained on 2022 UST data. The first changepoints are detected using the Markout metric on March 9th. All metrics emit changepoint alerts several hours before USDC dips below $99$ cents, with the earliest at 9pm UTC on March 10th. For reference, USDC first dips below $99$ cents at approximately 2am UTC on March 11th 2023 according to Chainlink oracles.}
        \label{fig:3pool_detected_cps}
    \end{figure}

    At a high level, our model performs particularly well using our entropy and markout metrics, especially for larger and more established pools such as the 3pool and ETH/stETH that don't exhibit as much noise. We find that our \textsc{bocd} model predicts the USDC depeg as early as 9pm UTC on March 10th, approximately five hours before USDC dips below $99$ cents according to Chainlink oracles, shown in Fig. \ref{fig:3pool_detected_cps}. That is, our models quickly captures structural changes in trading volume, pnl, and AMM composition before the overall market prices in a potential depeg. 
    
    However, for smaller pools such as USDN, sUSD, and stETH concentrated we obtain a lower precision score due to noisy metrics. We found that the PIN metric was not particularly informative, as discussed in Section \ref{sec:metrics}. Furthermore, we find that the Gini Coefficient, net LP flow, and Log Returns metrics are too noisy, and lead to the lowest precision and $lF$-scores as shown in Appendix \ref{app:fpr}. 
    
    We will focus on the following models in this discussion:
    
    \begin{enumerate}
        \item \textbf{Entropy model} using the hourly logarithmic differences on Shannon's entropy for each pool.
        \item \textbf{Markout model} using the hourly cumulative 5-minute markouts for each pool.
        \item \textbf{Swaps model} using the hourly cumulative swap flows for each pool.
    \end{enumerate}

    \subsection{Case Study - The 3pool}

        \begin{figure}[!htbp]
            \centering
            \includegraphics[width=\linewidth]{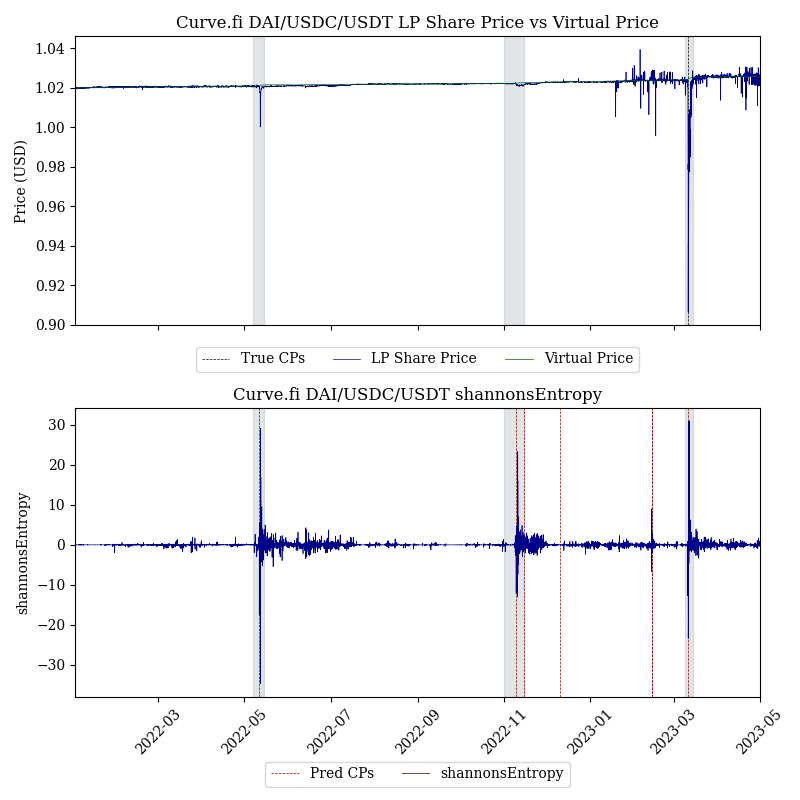}
            \caption{\textsc{bocd} results using hourly log differences in Shannon's Entropy for the 3pool. Our entropy model detects changepoints during the UST depeg, the FTX collapse, and the SVB bank run. The changepoint corresponding to the SVB collapse and the momentary depeg of USDC is detected at 9pm UTC on March 10th, hours before USDC dips below $99$ cents. Notice how increased variance in entropy closely correlates with all three high-information events, illustrated with the gray bars. Since 3pool LP share prices did not meaningfully deteriorate during the UST and FTX events, the corresponding changepoints are flagged as false positives by our scoring rule, although they are not necessarily ``false alarms'' in the pragmatic sense. We observe two additional false positives throughout late 2022 and early 2023.}
            \label{fig:3pool_entropy}
        \end{figure}

        The results for the 3pool were the most promising in our study, particularly using the Entropy model (Fig. \ref{fig:3pool_entropy}) and the Markout model. This is likely due to the 3pool being the largest pool on Curve, meaning individual swaps, deposits, and withdrawals have a relatively smaller effect on our observed metrics, leading to less noisy predictions. As hypothesized, most of our metrics are highly correlated with high information events, where the market is uncertain about the peg of tokens within the pool. For example, the collapse of UST led to market turmoil regarding USDT, one of the tokens in the 3pool, such that it traded as low as $95$ cents on May 12th \cite{bitscreener2022}. Accordingly, all our models detected changepoints on the 3pool between May 9th and May 11th, 2022.
        
        Conversely, the Gini Coefficient and Net LP Flow metrics are extremely noisy, with hundreds of false alarms throughout 2022 and 2023. Results for the 3pool are generally representative of results for the FRAX/USDC pool, also one of the largest pools on Curve, allowing us to attain higher precision in our models. Both markout and entropy have few false alarms, and detect a depeg during the SVB collapse at 10pm UTC on March 10th, 2023.

    \subsection{Case Study - ETH/stETH}

        \begin{figure}[!htbp]
            \centering
            \includegraphics[width=\linewidth]{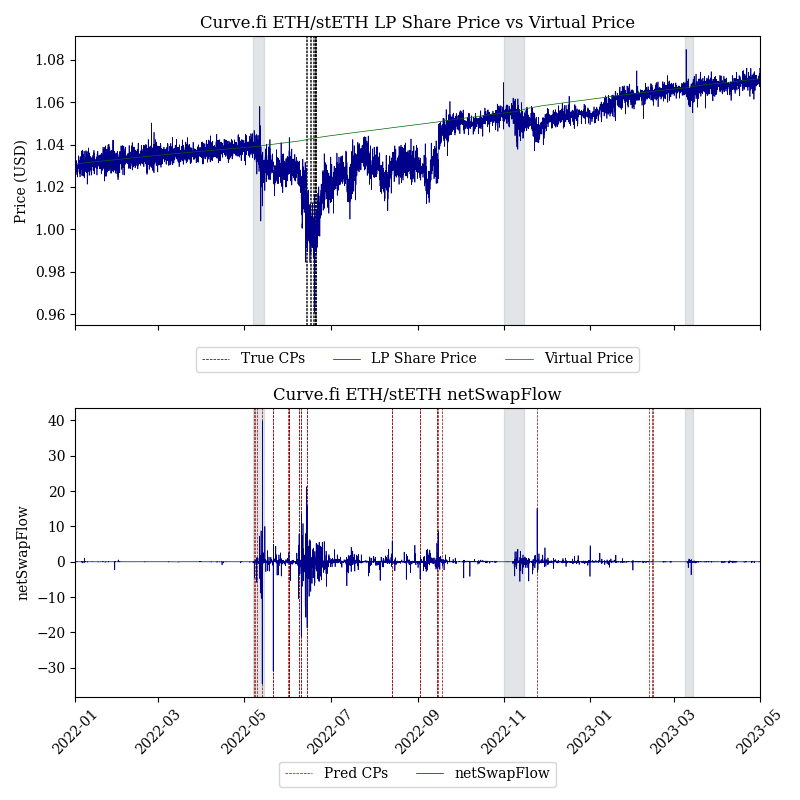}
            \caption{\textsc{bocd} results for the net swap flow metric in the ETH/stETH pool. Notice how the LP share price on the ETH/stETH pool is below its ``peg'' (the virtual price) for several months in 2022. Accordingly, most metrics, including the netSwapFlow metric pictured here, detect several changepoints as stETH depegs and repegs to ETH between June and October of 2022. stETH first depegs from ETH around May 12th 2022, and depegs further around June 10th. The first depeg detected by our Swaps model occurs on May 8th 2022, when stETH is still around $0.99$ ETH. The volatility during this period results in a number of depegs being detected by our models, although only the first detection provides useful leading information to liquidity providers.}
            \label{fig:stETH_swaps}
        \end{figure}

        Figure \ref{fig:stETH_swaps} depicts the results of our Swaps model on the ETH/stETH pool. Notice how the LP share price is much noisier for ETH/stETH than the 3pool, since stETH's peg to ETH is less stable than USDC, USDT, and DAI's peg to the dollar. Our metrics are slightly noisier with ETH/stETH than the 3pool, although they still detect changepoints hours before stETH prices fall below $0.99$ ETH. During the months for which stETH is below its peg, all our models detect several changepoints in the ETH/stETH pool, leading to low precision and high recall scores. 

        Because stETH's peg to ETH is fluctuates at a greater amplitude than stablecoins fluctuate relative to the dollar, the 5 minute markout model becomes much noisier. That is, we trained the hyperparameters for the Markout model using UST data, which was very closely pegged to the dollar before its collapse, leading to low markouts. The trained hyperparameters are then very sensitive to small fluctuations in markouts, which is useful for stablecoin pools but much noisier for LSD pools. In general, all our models are better suited to detect changepoints in assets that are tightly pegged to their numeraire, and have lower precisions when pegs are looser. As we will discuss in Section \ref{sec:takeaways}, this might be solvable by training LSD and Stablecoin pools separately, although there is not enough LSD depeg data to do so with high fidelity.
        
        Results for stETH concentrated are generally worse than ETH/stETH, especially since stETH concentrated was launched halfway through 2022, wherein it was particularly noisy. 

    \subsection{Case Study - USDN}

        \begin{figure}[!htbp]
            \centering
            \includegraphics[width=\linewidth]{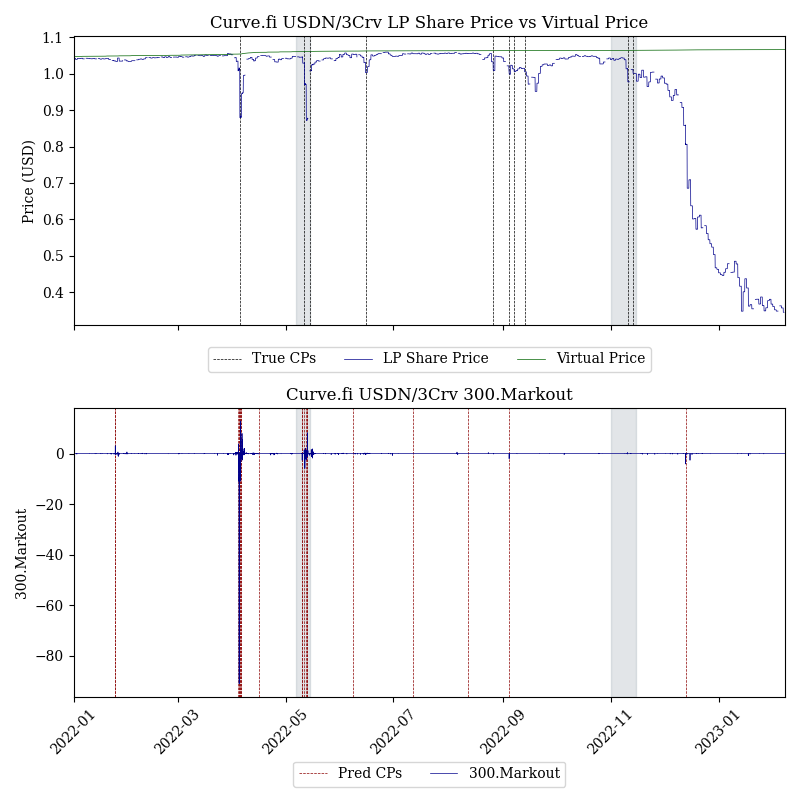}
            \caption{\textsc{bocd} results for the markout metric in the USDN Metapool. Markout changepoint detections generally coincide with changes in the LP share price relative to the virtual price, although none of the detections lead USDN price data. All models, including the markout model, become significantly noisier past June 2022, when the TVL and trading activity on the USDN pool declines significantly.}
            \label{fig:USDN_markout}
        \end{figure}

        The USDN pool was significantly smaller than other Curve pools such as the 3pool and UST Wormhole. The pool's TVL begins 2022 in the hundreds of millions of dollars, and falls to below $\$10M$ by late 2022. Our models do not detect changepoints for USDN in November of 2022, when it begins to gradually and permanently depeg from the dollar, likely due to the lack of liquidity and trading activity in the pool. Furthermore, the small amount of liquidity in the USDN pool relative to other Curve pools leads several of our models to be very noisy, especially those pertaining to pool composition such as entropy. 
        
        However, USDN kept a relatively tight peg to the dollar throughout most of 2022. Accordingly, our markout model is less noisy and detects depegs for USDN during both of its momentary depegs in April and May of 2022. However, the detections occur on April 4th and May 10th respectively, at which point USDN is already well below $99$ cents. Our models do not seem to be accurate leading indicators of depegs in these smaller pools, perhaps due to their training set, or due to the fact that trading data on Curve (such as swap amounts or pool composition) do not lead market prices as much as on the 3pool or ETH/stETH.

    \subsection{Case Study - Others}

        \begin{figure}[!htbp]
            \centering
            \includegraphics[width=\linewidth]{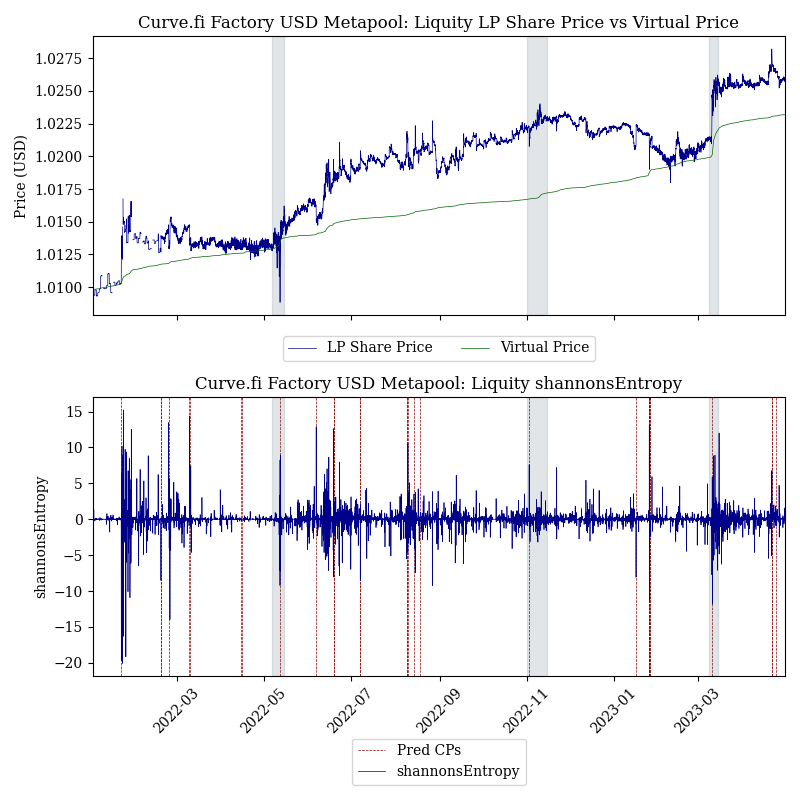}
            \caption{\textsc{bocd} results for the entropy metric in the Liquity USD Metapool. Notice how there are no ``true'' changepoints detected for the LUSD pool, as it is consistently 50 to 100bps above the virtual price and our threshold detection is $5\%$. As LUSD is a relatively small pool, individual swaps, deposits, and withdrawals have significant impact on pool composition, so both Entropy and Gini Coefficients are very noisy. Results for the LUSD pool generally represent the results on most smaller pools, such as the BUSD v1 metapool or the sUSD lending pool.}
            \label{fig:LUSD_entropy}
        \end{figure}

        Most other pools we have tested, aside from the 3pool, FRAX/USDC, ETH/stETH, and USDN, result in lower precision scores. That is, metrics for these pools tend to be noisier, where individual swaps, deposits, and withdrawals lead to large hour-over-hour differences despite no new information regarding token prices. Furthermore, the Log Returns model is very noisy for all tokens. This might be due to a generally noisy log returns metric, or due to the model overfitting to the UST log returns training data. 

    \subsection{Changepoint Intensity}

\begin{figure}[!htbp]
            \centering
            \includegraphics[width=\linewidth]{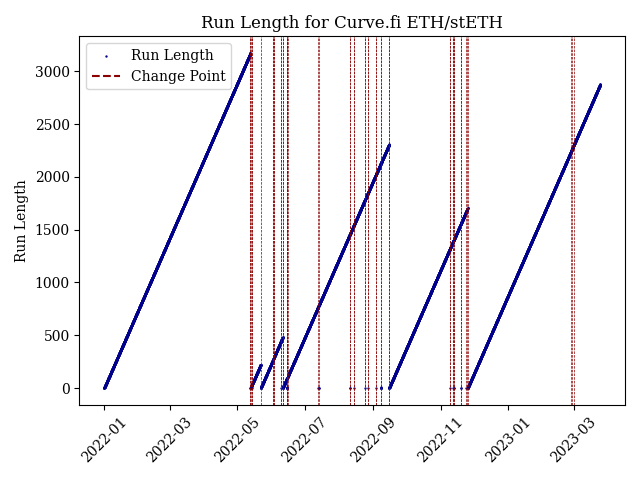}
            \caption{The \textsc{bocd} run length for the ETH stETH pool throughout the testing period. Notice that, while several changepoints were detected throughout the testing period, only 5 of them resulted in a persistent change to the model's run length.}
            \label{fig:run_length}
        \end{figure}

        Recall that in Section \ref{sec:detection} we defined a changepoint as any time $t$ where the run length $r_t \neq r_{t-1} + 1$. Notice this does not mean that the \textsc{bocd} model disconsiders the possibility of the run length extending on the next time step from $r_{t-1}$. That is, we might observe the following sequence: $[r_{t-1}, 0, r_{t-1} + 2]$, meaning the model mistakenly classified time $t$ as a changepoint, but corrected itself in the next time step. This might become clearer if we compare the run-length plot in Fig. \ref{fig:run_length} with Fig. \ref{fig:stETH_swaps}. While we do not incorporate this into our $lF$-score, it appears that many of the false positives detected by the model are quickly corrected for.

    \subsection{Conclusion}

        Our \textsc{bocd} model performs well on pools with large volume and large TVL, including the 3pool, ETH/stETH, and FRAX/USDC, whereas results are much noisier for smaller pools with less stable metrics. Furthermore, we find that the net swap flow, shannon's entropy, and 5-minute markout metrics are the most predictive metrics across all pools, with the Gini Coefficient, net deposits and withdrawals flow, and log returns being too noisy. 

        For large pools, we observe relatively few false positives, with most of our models detecting depegs several hours before prices dip below $99$ cents. However, as previously mentioned, there are very few historical depegs with which we can train and test our models. Results should be interpreted accordingly. Furthermore, a large portion of deposits into the 3pool originate in the various stablecoin metapools. Therefore, changepoints detected on the 3pool are relevant information for those LP'ing on the various metapools. Poor performance on the metapools themselves does not mean that we have no useful information for those LPs, since they may additionally listen to changepoints occuring on their base pool. 

        Generally, entropy is a strong metric to detect depegs on pools with large TVL, as individual swaps, deposits, and withdrawals have a smaller impact on the underlying metric, leading to higher precision scores. Similarly, net swap flows obtain higher precision scores for pools with large volume being processed, and markouts rely on prices being relatively stable around their peg in the period preceding a depeg. All three conditions hold for large stablecoin pools, whereas prices fluctuate more amply around their pegs on LSD pools, making markouts noisier.         

%% file: sections/takeaways.tex
\section{Takeaways} \label{sec:takeaways}
 
    In this paper we have presented several metrics for detecting depegs on Curve's stablecoins. We have designed and tested a Bayesian Online Changepoint Detection model using these metrics on 2022 and 2023 trading and pricing data, and identified 2-3 useful models for detecting depegs on Curve's largest StableSwap pools. We have shown that these detection models provide leading indicators of depegs by several hours, with few false positives in the 17 months on which they were tested. In this section, we describe how we believe this research may be leveraged by Curve LPs, and how we may extend our findings to better parameterize Curve pools.


        \subsection{Alerting Bot}

            We have deployed a Twitter bot for LPs to listen to alerts regarding the pools they LP into. We are making these services available for large StableSwap pools, and we believe they might be particularly helpful to non-systemic LPs who rebalance their portfolios on a discretionary basis. We aim to benefit Curve's tokenholders, who are oftentimes LPs on large Curve pools, by providing them with as much useful real-time information as possible to make informed decisions. Furthermore, we hope that sophisticated Curve users might find the results in this paper interesting, and perform further research and development using changepoint detection models for safer stablecoin liquidity provision. However, we do not recommend any LPs to naively integrate these alerts into their systems without performing their own backtesting. As previously stated, providing liquidity during potential depegs may be profitable for LPs given the increase in swap activity, and depends largely on the LP's risk appetite. 

        \subsection{Analytics}
    
        This research was sponsored by the Curve analytics team. All of our data, including metrics and detected changepoints, are available through our API. We hope this data will help develop and populate dashboards or other tools to inform Curve users on historical volatility for different pools, and highlight important historical depegs for further analysis. These metrics may be used in real-time in a variety of ways, not necessarily by implementing a Bayesian detection model. Furthermore, the real-time ranking of takers on Curve based on 1 day markouts are also available through our API, including their cumulative buys, sells, and trade count, as well as a query for identifying sharks.

        \subsection{Dynamic Parameter Updates}
        
            An exciting extension of the work presented in this paper is to identify opportunities for updating pool parameters in anticipation of asset depegs. This would allow the Curve DAO to dynamically derisk Curve pools for Curve LPs and mitigate their losses. This could be done, for example, by lowering the $A$ parameter in the pool, effectively increasing the constant-product element of the pool which increases the cost of swapping a depegging token for a pegged token. Similarly, we might raise fees in anticipation of an asset depeg in a StableSwap pool. In Curve v1, neither is done automatically: both the $A$ parameter and fees are constant and may only be updated through governance proposals. In Curve v2, fees are dynamic, and change according to the imbalance of pool reserves\footnote{For an in-depth explanation of Curve's v2 parameters, refer to this post by Nagaking \cite{nagaking2022deep}.}. Here we describe some preliminary thoughts for how we may set up an entity responsible for listening to depeg alerts, running relevant simulations, and dynamically updating pool parameters with minimal governance overhead.

            To understand how lowering the $A$ parameter might de-risk a StableSwap pool for LPs if any constituent asset risks momentarily or permanently depegging we must refer back to the StableSwap invariant. Let us first parameterize the invariant using the ``leverage'' parameter $\chi$.

            \begin{equation} \label{eq:invariant_chi}
                \chi D^{n-1} \sum{x_i} + \Pi x_i = \chi D^n + \left(\frac{D}{n}\right)^n
            \end{equation}

            where $x_i$ are token balances, $D$ is the sum of token balances at the equilibrium point, $n$ is the number of tokens, and $\chi$ is the leverage parameter, defined below.
        
            \begin{equation} \label{eq:chi}
                \chi = \frac{A \Pi x_i}{(D/n)^n}
            \end{equation}

            where $A$ is the amplification coefficient, and determines the portion of the StableSwap AMM that is determined by the constant-sum AMM, $\chi D^{n-1} \sum{x_i}$, versus the constant product AMM $\Pi x_i$. Substituting Eq. \ref{eq:chi} in Eq. \ref{eq:invariant_chi} leads us to the canonical StableSwap invariant:

            \begin{equation}
                An^n\sum{x_i} + D = ADn^n + \frac{D^{n+1}{n^n \Pi x_i}}.
            \end{equation}

            It might now be clear why lowering $A$ leads to a constant-price AMM: when $A=0$, $\chi=0$, so Eq. \ref{eq:invariant_chi} becomes a constant price AMM. That is, in a constant product AMM, the marginal deviation from equilibrium (i.e. any swap) incurs a greater cost $\frac{\partial x_i}{\partial x_j}$ than in a constant sum AMM. Therefore, by lowering $A$, we increase the marginal slippage incurred by a trader swapping out of a depegging asset at all price points, including at equilibrium. If we expect a token to depeg, this might be desirable for LPs, as it would dampen the rate at which LPs sell their pegged tokens and buy the depegging token. This makes the pool a more attractive investment for LPs when token pegs are uncertain, and gives them more time to pull out of the pool before their balances are significantly affected by depeg trading.

            A possible first step would be to use simulations, such as \href{https://github.com/curveresearch/curvesim}{\texttt{curvesim}} to model different parameters (i.e. $A$ and fees) during depegs (e.g. UST) and see if changes to these parameters would create clear benefits to LPs. If so, then we could backtest making these changes when our models emitted an alert and observe the effect on LP profitability. We could further leverage the metrics and models developed herein to identify the most useful periods in time to model using \href{https://github.com/curveresearch/curvesim}{\texttt{curvesim}}. For example, we might search for when our entropy metric is most volatile for a particular pool, and model changes to the amplification coefficient during that period.

            A concern is that depegs were detected in at most $5$ hours preceding the first price drops. Furthermore, these depegs were often detected by an entropy metric, which itself relies on changes to the pool's composition. At this point, it might be too late to enact parameter changes for two reasons:

            \begin{enumerate}
                \item Making changes to $A$ while the pool is imbalanced might lead to losses in $D$. For an example, refer to \href{https://gov.curve.fi/t/proposal-to-lower-a-parameter-of-peth-eth-pool-10-6/9008}{this} goverance proposal. This means that lowering $A$ based on changes to entropy might not be prudent, as it would lead to losses with respect to the pool's virtual price.
                \item Enacting parameter changes requires a governance vote, which itself is a lengthy process. By the time the vote is executed, the depeg might already have occured.
            \end{enumerate}

            The first concern might be ameliorated by considering other non-pool-composition-based metrics, such as net swap flow or markouts, or by increasing fees without changing $A$.
                The second concern may be addressed in a few ways: by a subDAO committee that manages pool parameters, by a third-party vault application on top of Curve, or by systematic on-chain verifiable parameter updates.
    
                One possibility would be to nominate a subDAO to control pool parameter values, however this would be operationally burdensome and potentially harmful to Curve's decentralization. Another option would be to create a vault on top of Curve that actively manages Curve LP positions to remove liquidity when changepoints are detected, though this has the downside of not integrating directly into Curve, thus requiring LPs to interface with the third-party vault protocol.
                
                Another option for solving concern (2) would be to create a fully systematic parameter maintainer. That is, we would use on-chain data -- such as reserves, swap flow, LP withdrawal events -- to construct metrics, then determine if a changepoint has occurred in those metrics, then push an update to the pool's $A$ and/or fee parameters based on the changepoint detection. Of course, running this calculation on-chain is intractable, but we could instead generate a proof (e.g., a SNARK) of the calculation off-chain, submit the proof for verification on-chain, then make the pool parameter changes on-chain if the proof is valid. Such behavior would enable us to systematically make pool changes. Undeniably, this approach has concerns of its own, most notably the development complexity and the dependence on our changepoint metrics.
                
                We believe that dynamic parameter updates are one of the most exciting applications of our research, as they have the potential to make Curve an even better venue for passive LPs. There are a number of ways to achieve this, each with its own tradeoffs, but we are confident that Curve's LPs stand to benefit from additional research and development on this topic.

    \subsection{Discussion} \label{subsec:discussion}

        Here we describe some potential extensions to our work.

        \subsubsection{Metric Ensemble Methods}

            Each of our \textsc{bocd} models was considered independently. We did not consider building an ensemble model on top of a subset of our \textsc{bocd} models that aggregates their outputs to determine the probability of a potential depeg. An example might be a simple decision tree, that takes as inputs the outputs of each of our models, and determines whether enough models have recently detected a depeg in order to fire an alert to LPs. Notice that each of the predicted changepoints from our \textsc{bocd} models come with a corresponding probability (i.e. $P(r_t)$), which may further be used as an input to the ensemble model. Given that model predictions have an associated probability of the run length being $0$, general model calibration techniques might be fruitful in reducing the number of false positives for each of our models.



        \subsubsection{More Data}

            A straightforward enhancement to our work is to train and test our models on more data. Particularly, we might gather data on depegs farther back in time, as well as consider additional pools and tokens in our analysis. 

        \subsubsection{Additional Metrics}

            We have developed metrics based primarily on price data and Curve trading data. We could imagine developing similar metrics using trading data on other venues, or other novel metrics. For example, we may consider:
            
            \begin{itemize}
                \item A net swap flow metric aggregated over Uniswap, Curve, and Balancer
                \item A sentiment analysis study for each token, where negative sentiments indicates depegs. We may perform sentiment analysis on Twitter for a particular token, and perform our Bayesian Online Changepoint Detection (or other detection algorithm) on the output sentiments.
                \item An implied volatility index for stablecoins and LSDs would likely be the best leading indicator of token depegs. However, due to the lack of depth in options markets for stablecoins, or crypto currencies in general, such implied volatility indices do not exist.
                \item VPIN - an adaptation of the PIN metric that uses volumetric time instead of standard time in measuring market asymmetry. Given that a large concern with the PIN metric is that it doesn't consider volume (only net amount of trades), VPIN might provide more reliable results.
            \end{itemize}

    \subsection{Model Deployment} \label{subsec:deployment}

        We find that \textsc{bocd} is sensitive to hyperparameters, which means that its performance might degrade over time if hyperparameters are not actively adjusted. This sensitivity is largely a function of the number of depegs contained in the training set. If no depegs are available, then there will be no labels on which to train our models. For this reason, training \textsc{bocd} against arbitrary periods, such as training a model using randomly sampled months or years of data, yields poor results.
        
        We have deployed \textsc{bocd} using the hyperparameters that we have tested, i.e. those trained on the UST depeg, which have performed well against a large testing set. We have deployed models for the Shannon's Entropy, Net Swap Flow, and 5-minute Markout metrics for the larger StableSwap pools on Curve. You may find more information in our repo's \texttt{README}.

%% file: appendix/bocd_setup.tex
\section{\textsc{BOCD} Setup} \label{app:bocd_setup}

    Bayesian online changepoint detection is composed of two primary components:

    \begin{enumerate}
        \item A hazard function $H$
        \item A likelihood function $\pi$
    \end{enumerate}

    Here, we discuss why we chose a constant hazard function and a Student's t-distribution as our likelihood function. We further describe how we performed the hypermarater tuning for the parameters involved in each component.

    \subsection{Hazard Function}

        The hazard function answers the following: given a run length $r_t$, what is the probability that the next data point is a change point, given no other information? For example, if we are detecting changepoints to the output of some machine, and we know that after $N$ executions the machine begins to degrade in quality, we might want to input this degradation into \textsc{bocd} through the hazard function.

        With respect to asset depegs, we find no obvious run-length-dependency with respect to changepoints; depegs do not necessarily become more or less likely over time. For this reason, we use the following constant hazard function:

        \begin{equation}
            H(r_t) = \frac{1}{\lambda}
        \end{equation}

        where we set $\lambda=100$ in all our model runs.

    \subsection{Likelihood Function}

        The likelihood function is the probability density function (pdf) of a statistical distribution, denoted as $\pi$. The pdf determines the probability that a new observation $x$ belongs to the previously observed distribution. The notation $\pi_{t-1}^{l}$ denotes the pdf parameterized at time $t-1$ with run-length $l$. With every new \textsc{bocd} update, we update the parameters in the pdf according to the new $l$ observations at time $t-1$.
    
        The likelihood function allows us to include some prior information about our data into our \textsc{bocd} model. For example, if we are modeling the amount of car crashes on some interstate, we might know that they arrive as some Poisson point process. Using historical data, we create some prior distribution for this Poisson point process to act as the base case in our \textsc{bocd} model (i.e. when $r_t = 0$), and with every new datum $x_t$ we update the parameters in our Poisson distribution. In this case, the pdf of the Poisson distribution is the likelihood function that determines if the number of car crashes observed during a particular period of time is unlikely to have been sampled from our prior distribution.

        We make a simplifying assumption in this paper that all our metrics can be transformed into being approximately Gaussian, for example by taking the log differences. However, since outliers in our data (i.e. 3+ standard deviation observations) occur relatively frequently in our data, and those are precisely the data points we want to detect, we use a Student's t-distribution to control the tail size of our prior distribution. More specifically, the t-distribution lets us control the kurtosis of the distribution we use to model each metric. 
        
        We use Normal-Gamma parameters to parameterize our t-distribution:

        \begin{equation}
            t(x; \mu, \alpha, \beta, \kappa) = 
        \frac{\Gamma((2\alpha+1)/2)}{\sqrt{\beta(2\alpha)\pi}\Gamma(\alpha)} 
        \left(1 + \frac{\kappa(x - \mu)^2}{2\alpha\beta}\right)^{-((2\alpha+1)/2)}
        \end{equation}

        where $\Gamma$ is the Gamma function, and $\mu, \alpha, \beta, \kappa$ are Normal-Gamma parameters which can be mapped onto the usual Student's t parameters - mean, degrees of freedom, and scale - as follows:

        \begin{align}
            \mu = \mu \\
            \sigma^2 = \frac{\beta}{\alpha \kappa} \\
            \nu = 2\alpha
        \end{align}

        The $\alpha, \beta, \kappa$ parameters control how sensitive our \textsc{bocd} model will be to the variance in our observations. To build some intuition, we may imagine that a larger $\kappa$ decreases the variance $\sigma$, which in term makes our distribution more concentrated around the mean $\mu$. This means we are more certain that observations will tightly fluctate around the mean, such that slight deviations from $\mu=0$ might be flagged as changepoints. Higher $\beta$ makes our distribution flatter, such that \textsc{bocd} is less sensitive. Similarly, higher $\alpha$ increases the degrees of freedom parameter $\nu$ in our distribution, which affects the shape of the tails (i.e. makes it more or less similar to a Gaussian distribution). Larger $\alpha$ increases the kurtosis of our distribution, making the model less sensitive to outliers.
        
        Using Normal-Gamma parameters for our Student's t-distribution allows us to leverage conjugacy between the prior and posterior of our Bayesian updates, making computation much simpler \cite{murphy2007}. These are the hyperparameters that we train for each metric.
    
    \subsection{Hyperparameter Tuning}

        Hyperparameter tuning (i.e. model training) is performed on 6 months of data on the UST Wormhole Metapool. This was chosen specifically because it includes 1 month of data on a relevant depeg, as well as 5 months of data on non-depegging data, creating a somewhat balanced training set. Furthermore, by including only one depeg in our training set, we are able to observe model performance on a variety of other depegs, including USDC, USDN, and stETH.

        As mentioned, we do not perform hyperparameter tuning on the $\lambda$ parameter of the constant Hazard function, nor do we tune the $\mu=0$ mean for the Student's t-distribution. For $\alpha, \beta, \kappa$, we grid search optimal configurations over the following parameter space:

        \begin{equation}
            \mathcal{P} = \{[\alpha^i, \beta^j, \kappa^k]\}, \forall i, j, k \in [-5, \ldots, 4]
        \end{equation}

        For each metric and configuration $[\alpha^i, \beta^j, \kappa^k]$ we run our \textsc{bocd} model and compute its corresponding $lF$-score. We then select the configuration with the highest $lF$-score for each metric. 

%% file: appendix/pools.tex
\section{Supported Pools} \label{app:pools}

    The pools supported in this paper and on our API are displayed in Table \ref{table:pools}, along with their respective Ethereum addresses. Note, we only support Ethereum pools.

    \begin{center}
        \captionof{table}{Supported Pools \label{table:pools}}
        \begin{tabular}[\textwidth]{l r}
            \toprule
            Pool & Address \\
            \midrule
            3pool & \texttt{0xbebc44782c7db0a1a60cb6fe97d0b483032ff1c7} \\
            steth & \texttt{0xdc24316b9ae028f1497c275eb9192a3ea0f67022} \\
            fraxusdc & \texttt{0xdcef968d416a41cdac0ed8702fac8128a64241a2} \\
            UST wormhole & \texttt{0xceaf7747579696a2f0bb206a14210e3c9e6fb269} \\
            USDN & \texttt{0x0f9cb53ebe405d49a0bbdbd291a65ff571bc83e1} \\
            mim & \texttt{0x5a6a4d54456819380173272a5e8e9b9904bdf41b} \\
            susd & \texttt{0xa5407eae9ba41422680e2e00537571bcc53efbfd} \\
            frxeth & \texttt{0xa1f8a6807c402e4a15ef4eba36528a3fed24e577} \\
            lusd & \texttt{0xed279fdd11ca84beef15af5d39bb4d4bee23f0ca} \\
            busdv2 & \texttt{0x4807862aa8b2bf68830e4c8dc86d0e9a998e085a} \\
            stETH concentrated & \texttt{0x828b154032950c8ff7cf8085d841723db2696056} \\
            cbETH ETH & \texttt{0x5fae7e604fc3e24fd43a72867cebac94c65b404a} \\
            cvxCRV CRV & \texttt{0x971add32ea87f10bd192671630be3be8a11b8623} \\
            \bottomrule
        \end{tabular}
    \end{center}

%% file: appendix/sources.tex
\section{Price Data Sources} \label{app:sources}

    The prices we use for each token are displayed in Listing \ref{lst:sources}. We prefer \texttt{ccxt} prices for convenience given \texttt{ccxt} data is more regular, but cross-check with Chainlink oracle feeds to confirm the accuracy of our data.
    
    \begin{center}
        \captionof{listing}{Price Data Sources \label{lst:sources}} 
        \begin{minted}[frame=single,
                       framesep=3mm,
                       linenos=true,
                       xleftmargin=21pt,
                       tabsize=4]{js}
        {     
            "token_exchange_map" : {
                "USDC" : ["ccxt", "binanceus"],
                "MIM" : ["ccxt", "bitfinex2"],
                "UST" : ["ccxt", "coinbasepro"],
                "cbETH" : ["ccxt", "coinbasepro"],
                "ETH" : ["ccxt", "binanceus"],
                "USDT" : ["ccxt", "binanceus"],
                "CRV" : ["ccxt", "binanceus"],
                "DAI" : ["ccxt", "binanceus"],
                "FRAX" : ["chainlink", "0xB9E1E3A9feFf48998E45Fa90847ed4D467E8BcfD"],
                "BUSD" : ["chainlink", "0x833D8Eb16D306ed1FbB5D7A2E019e106B960965A"],
                "LUSD" : ["chainlink", "0x3D7aE7E594f2f2091Ad8798313450130d0Aba3a0"],
                "sUSD" : ["chainlink", "0xad35Bd71b9aFE6e4bDc266B345c198eaDEf9Ad94"],
                "stETH" : ["chainlink", "0xCfE54B5cD566aB89272946F602D76Ea879CAb4a8"],
                "USDN" : ["chainlink", "0x099c9588D8C6F7579C89014e59002881CE0c46A1"]
            }
        }
        \end{minted}
    \end{center}

%% file: appendix/fpr.tex
\section{\textsc{bocd} Results} \label{app:fpr}

    Table \ref{table:pool_results} summarizes the results of our \textsc{bocd} tests for each pool and metric pair. The $lF$-scores are generally \textit{much} lower than usual $F$-scores for machine learning models. There are two reasons for this:

    \begin{enumerate}
        \item We use a weighted recall metric, meaning most true positives have a weight less than $1$. This is in contrast to regular recall metrics, which weigh all true positives at $1$. This leads to low recall, despite accurately predicting every depeg in the testing set for that pool. 
        \item Our labels generally correspond to the first period at which LP share prices drop below $5\%$ of the virtual price. This means that the first detected depeg will ``match'' to that label, and any changepoints detected shortly thereafter will be labelled as false positives. Ideally, we would only detect $1$ changepoint before prices depreciate. Of course, the moments following a depeg are often very volatile, which is why in practice we observe short bursts in detected changepoints. This leads to low precision scores.
    \end{enumerate}

    Since we do not use the $lF$-score to benchmark our model against other models, we merely use them for hyperparameter tuning, these distortions relative to standard F-scores are acceptable.

    \begin{center}
        \tiny
        \captionof{table}{Pool Results \label{table:pool_results}}
        \begin{tabular}{llrrrrrr}
            \toprule
             &  & F & P & R & alpha & beta & kappa \\
            pool & metric &  &  &  &  &  &  \\
            \midrule
            \multirow[t]{6}{*}{\texttt{0xdc24316b9ae028f1497c275eb9192a3ea0f67022}} & shannonsEntropy & 0.02758 & 0.01754 & 0.06439 & 0.10000 & 1000 & 1.00000 \\
             & sharkflow & 0.06684 & 0.03620 & 0.43561 & 0.00001 & 1 & 10000.00000 \\
             & giniCoefficient & 0.00000 & 0.00000 & 0.00000 & 100.00000 & 100 & 10000.00000 \\
             & netLPFlow & 0.02336 & 0.01471 & 0.05682 & 0.10000 & 100 & 1000.00000 \\
             & netSwapFlow & 0.02128 & 0.03571 & 0.01515 & 0.01000 & 1000 & 1.00000 \\
             & 300.Markout & 0.00000 & 0.01333 & 0.00000 & 10.00000 & 100 & 0.00010 \\
            
            \multirow[t]{6}{*}{\texttt{0xbebc44782c7db0a1a60cb6fe97d0b483032ff1c7}} & shannonsEntropy & 0.21277 & 0.14286 & 0.41667 & 0.10000 & 1000 & 1.00000 \\
             & sharkflow & 0.02277 & 0.01163 & 0.54167 & 0.00001 & 1 & 10000.00000 \\
             & giniCoefficient & 0.00984 & 0.00495 & 0.75000 & 100.00000 & 100 & 10000.00000 \\
             & netLPFlow & 0.01700 & 0.00870 & 0.37500 & 0.10000 & 100 & 1000.00000 \\
             & netSwapFlow & 0.11111 & 0.07143 & 0.25000 & 0.01000 & 1000 & 1.00000 \\
             & 300.Markout & 0.05650 & 0.03030 & 0.41667 & 10.00000 & 100 & 0.00010 \\
            
            \multirow[t]{6}{*}{\texttt{0xdcef968d416a41cdac0ed8702fac8128a64241a2}} & shannonsEntropy & 0.14286 & 0.12500 & 0.16667 & 0.10000 & 1000 & 1.00000 \\
             & sharkflow & 0.06780 & 0.03774 & 0.33333 & 0.00001 & 1 & 10000.00000 \\
             & giniCoefficient & 0.05882 & 0.03226 & 0.33333 & 100.00000 & 100 & 10000.00000 \\
             & netLPFlow & 0.04138 & 0.02326 & 0.18750 & 0.10000 & 100 & 1000.00000 \\
             & netSwapFlow & 0.00000 & 0.00000 & 0.00000 & 0.01000 & 1000 & 1.00000 \\
             & 300.Markout & 0.26496 & 0.16667 & 0.64583 & 10.00000 & 100 & 0.00010 \\
            
            \multirow[t]{6}{*}{\texttt{0x0f9cb53ebe405d49a0bbdbd291a65ff571bc83e1}} & shannonsEntropy & 0.01653 & 0.03030 & 0.01136 & 0.10000 & 1000 & 1.00000 \\
             & sharkflow & 0.04852 & 0.02643 & 0.29545 & 0.00001 & 1 & 10000.00000 \\
             & giniCoefficient & 0.02405 & 0.01370 & 0.09848 & 100.00000 & 100 & 10000.00000 \\
             & netLPFlow & 0.05714 & 0.03614 & 0.13636 & 0.10000 & 100 & 1000.00000 \\
             & netSwapFlow & 0.09231 & 0.14286 & 0.06818 & 0.01000 & 1000 & 1.00000 \\
             & 300.Markout & 0.08439 & 0.09524 & 0.07576 & 10.00000 & 100 & 0.00010 \\
            
            \multirow[t]{6}{*}{\texttt{0x5a6a4d54456819380173272a5e8e9b9904bdf41b}} & shannonsEntropy & 0.04152 & 0.06452 & 0.03061 & 0.10000 & 1000 & 1.00000 \\
             & sharkflow & 0.06494 & 0.07143 & 0.05952 & 0.00001 & 1 & 10000.00000 \\
             & giniCoefficient & 0.02159 & 0.02410 & 0.01956 & 100.00000 & 100 & 10000.00000 \\
             & netLPFlow & 0.03693 & 0.04167 & 0.03316 & 0.10000 & 100 & 1000.00000 \\
             & netSwapFlow & 0.02339 & 0.08333 & 0.01361 & 0.01000 & 1000 & 1.00000 \\
             & 300.Markout & 0.15199 & 0.20755 & 0.11990 & 10.00000 & 100 & 0.00010 \\
            
            \multirow[t]{6}{*}{\texttt{0xa5407eae9ba41422680e2e00537571bcc53efbfd}} & shannonsEntropy & 0.08451 & 0.04762 & 0.37500 & 0.10000 & 1000 & 1.00000 \\
             & sharkflow & 0.01095 & 0.00556 & 0.37500 & 0.00001 & 1 & 10000.00000 \\
             & giniCoefficient & 0.00000 & 0.00000 & 0.00000 & 100.00000 & 100 & 10000.00000 \\
             & netLPFlow & 0.01378 & 0.00699 & 0.45833 & 0.10000 & 100 & 1000.00000 \\
             & netSwapFlow & 0.08333 & 0.04762 & 0.33333 & 0.01000 & 1000 & 1.00000 \\
             & 300.Markout & 0.01709 & 0.00885 & 0.25000 & 10.00000 & 100 & 0.00010 \\
            
            \multirow[t]{6}{*}{\texttt{0xa1f8a6807c402e4a15ef4eba36528a3fed24e577}} & shannonsEntropy & 0.00000 & 0.00000 & 0.00000 & 0.10000 & 1000 & 1.00000 \\
             & sharkflow & 0.00000 & 0.00000 & 0.00000 & 0.00001 & 1 & 10000.00000 \\
             & giniCoefficient & 0.00000 & 0.00000 & 0.00000 & 100.00000 & 100 & 10000.00000 \\
             & netLPFlow & 0.00000 & 0.00000 & 0.00000 & 0.10000 & 100 & 1000.00000 \\
             & netSwapFlow & 0.00000 & 0.00000 & 0.00000 & 0.01000 & 1000 & 1.00000 \\
             & 300.Markout & 0.00000 & 0.00000 & 0.00000 & 10.00000 & 100 & 0.00010 \\
            
            \multirow[t]{6}{*}{\texttt{0xed279fdd11ca84beef15af5d39bb4d4bee23f0ca}} & shannonsEntropy & 0.00000 & 0.00000 & 0.00000 & 0.10000 & 1000 & 1.00000 \\
             & sharkflow & 0.00000 & 0.00000 & 0.00000 & 0.00001 & 1 & 10000.00000 \\
             & giniCoefficient & 0.00000 & 0.00000 & 0.00000 & 100.00000 & 100 & 10000.00000 \\
             & netLPFlow & 0.00000 & 0.00000 & 0.00000 & 0.10000 & 100 & 1000.00000 \\
             & netSwapFlow & 0.00000 & 0.00000 & 0.00000 & 0.01000 & 1000 & 1.00000 \\
             & 300.Markout & 0.00000 & 0.00000 & 0.00000 & 10.00000 & 100 & 0.00010 \\
            
            \multirow[t]{6}{*}{\texttt{0x4807862aa8b2bf68830e4c8dc86d0e9a998e085a}} & shannonsEntropy & 0.00000 & 0.00000 & 0.00000 & 0.10000 & 1000 & 1.00000 \\
             & sharkflow & 0.00000 & 0.00000 & 0.00000 & 0.00001 & 1 & 10000.00000 \\
             & giniCoefficient & 0.00000 & 0.00000 & 0.00000 & 100.00000 & 100 & 10000.00000 \\
             & netLPFlow & 0.00000 & 0.00000 & 0.00000 & 0.10000 & 100 & 1000.00000 \\
             & netSwapFlow & 0.00000 & 0.00000 & 0.00000 & 0.01000 & 1000 & 1.00000 \\
             & 300.Markout & 0.00000 & 0.00000 & 0.00000 & 10.00000 & 100 & 0.00010 \\
            
            \multirow[t]{6}{*}{\texttt{0x828b154032950c8ff7cf8085d841723db2696056}} & shannonsEntropy & 0.00000 & 0.00000 & 0.00000 & 0.10000 & 1000 & 1.00000 \\
             & sharkflow & 0.22398 & 0.17797 & 0.30208 & 0.00001 & 1 & 10000.00000 \\
             & giniCoefficient & 0.00000 & 0.00000 & 0.00000 & 100.00000 & 100 & 10000.00000 \\
             & netLPFlow & 0.16279 & 0.23333 & 0.12500 & 0.10000 & 100 & 1000.00000 \\
             & netSwapFlow & 0.13751 & 0.60000 & 0.07765 & 0.01000 & 1000 & 1.00000 \\
             & 300.Markout & 0.07629 & 0.08000 & 0.07292 & 10.00000 & 100 & 0.00010 \\
            
            \multirow[t]{6}{*}{\texttt{0x5fae7e604fc3e24fd43a72867cebac94c65b404a}} & shannonsEntropy & 0.00000 & 0.00000 & 0.00000 & 0.10000 & 1000 & 1.00000 \\
             & sharkflow & 0.00000 & 0.00000 & 0.00000 & 0.00001 & 1 & 10000.00000 \\
             & giniCoefficient & 0.00000 & 0.00000 & 0.00000 & 100.00000 & 100 & 10000.00000 \\
             & netLPFlow & 0.00000 & 0.00000 & 0.00000 & 0.10000 & 100 & 1000.00000 \\
             & netSwapFlow & 0.00000 & 0.00000 & 0.00000 & 0.01000 & 1000 & 1.00000 \\
             & 300.Markout & 0.00000 & 0.00000 & 0.00000 & 10.00000 & 100 & 0.00010 \\
            
            \multirow[t]{6}{*}{\texttt{0x971add32ea87f10bd192671630be3be8a11b8623}} & shannonsEntropy & 0.00000 & 0.00000 & 0.00000 & 0.10000 & 1000 & 1.00000 \\
             & sharkflow & 0.00000 & 0.00000 & 0.00000 & 0.00001 & 1 & 10000.00000 \\
             & giniCoefficient & 0.00000 & 0.00000 & 0.00000 & 100.00000 & 100 & 10000.00000 \\
             & netLPFlow & 0.00000 & 0.00000 & 0.00000 & 0.10000 & 100 & 1000.00000 \\
             & netSwapFlow & 0.00000 & 0.00000 & 0.00000 & 0.01000 & 1000 & 1.00000 \\
             & 300.Markout & 0.00000 & 0.00000 & 0.00000 & 10.00000 & 100 & 0.00010 \\
            
            \bottomrule
        \end{tabular}
    \end{center}

    \clearpage

    Table \ref{table:token_results} summarizes the results for the Log Returns model for the tokens contained within the pools in this study.

    \begin{center}
        \captionof{table}{Token Results \label{table:token_results}}
        \begin{tabular}{llrrrrrr}
            \toprule
             &  & F & P & R & alpha & beta & kappa \\
            token & metric &  &  &  &  &  &  \\
            \midrule
            \texttt{0x5f98805a4e8be255a32880fdec7f6728c6568ba0} & logReturns & 0.01 & 0.01 & 0.80 & 0.00 & 0.10 & 1000 \\
            
            \texttt{0x853d955acef822db058eb8505911ed77f175b99e} & logReturns & 0.00 & 0.00 & 0.90 & 0.00 & 0.10 & 1000 \\
            
            \texttt{0x99d8a9c45b2eca8864373a26d1459e3dff1e17f3} & logReturns & 0.14 & 0.08 & 0.58 & 0.00 & 0.10 & 1000 \\
            
            \texttt{0xa0b86991c6218b36c1d19d4a2e9eb0ce3606eb48} & logReturns & 0.01 & 0.00 & 0.52 & 0.00 & 0.10 & 1000 \\
            
            \texttt{0xae7ab96520de3a18e5e111b5eaab095312d7fe84} & logReturns & 0.00 & 0.00 & 0.00 & 0.00 & 0.10 & 1000 \\
            
            \texttt{0xbe9895146f7af43049ca1c1ae358b0541ea49704} & logReturns & 0.00 & 0.00 & 0.00 & 0.00 & 0.10 & 1000 \\
            
            \texttt{0x6b175474e89094c44da98b954eedeac495271d0f} & logReturns & 0.01 & 0.00 & 0.53 & 0.00 & 0.10 & 1000 \\
            
            \texttt{0x4fabb145d64652a948d72533023f6e7a623c7c53} & logReturns & 0.00 & 0.00 & 0.00 & 0.00 & 0.10 & 1000 \\
            
            \texttt{0x57ab1ec28d129707052df4df418d58a2d46d5f51} & logReturns & 0.00 & 0.00 & 0.00 & 0.00 & 0.10 & 1000 \\
            
            \texttt{0x62b9c7356a2dc64a1969e19c23e4f579f9810aa7} & logReturns & 0.29 & 0.21 & 0.46 & 0.00 & 0.10 & 1000 \\
            
            \texttt{0xdac17f958d2ee523a2206206994597c13d831ec7} & logReturns & 0.00 & 0.00 & 0.00 & 0.00 & 0.10 & 1000 \\
            
            \texttt{0x674c6ad92fd080e4004b2312b45f796a192d27a0} & logReturns & 0.04 & 0.02 & 0.99 & 0.00 & 0.10 & 1000 \\
            
            \texttt{0x5e8422345238f34275888049021821e8e08caa1f} & logReturns & 0.00 & 0.00 & 0.00 & 0.00 & 0.10 & 1000 \\
            
            \texttt{0xd533a949740bb3306d119cc777fa900ba034cd52} & logReturns & 0.09 & 0.05 & 0.78 & 0.00 & 0.10 & 1000 \\
            
            \bottomrule
        \end{tabular}
    \end{center}